\def\terseversion{1} 

\ifdefined\conciseversion
  \documentclass[12pt]{article}
\else
  \documentclass[12pt]{article}
\fi
\usepackage[T1]{fontenc}
\usepackage{lmodern}
\usepackage{microtype}
\usepackage{float}
\usepackage[utf8]{inputenc}
\usepackage{amsthm,amsmath}
\usepackage{url}
\usepackage{tabularx}
\usepackage{longtable}
\usepackage{array}
\usepackage{xcolor}
\usepackage{booktabs}
\usepackage{amsfonts}
\usepackage{comment}
\usepackage[hidelinks]{hyperref}
\usepackage[sort&compress,nameinlink,noabbrev,capitalize]{cleveref}
\usepackage{graphicx}
\usepackage[affil-it]{authblk}
\ifdefined\conciseversion
  \usepackage[margin=1in]{geometry}
\fi
\newif\ifconcise
\ifdefined\conciseversion \concisetrue \else \concisefalse \fi

\graphicspath{{figures/}{./}}
\newtheorem{remark}{Remark}

\title{Legitimate Overrides in Decentralized Protocols}
\author[1]{Oghenekaro Elem}
\author[2]{Nimrod Talmon}
\affil[1]{Parametrig, \href{mailto:karo@parametrig.com}{karo@parametrig.com}}
\affil[2]{BGU, IOG, \href{mailto:talmonn@bgu.ac.il}{talmonn@bgu.ac.il}}
\date{}
\begin{document}
\maketitle

\begin{abstract}
Decentralized protocols claim immutable, rule-based execution, yet many embed emergency mechanisms such as chain-level freezes, protocol pauses, and account quarantines. These overrides are crucial for responding to exploits and systemic failures, but they expose a core tension: when does intervention preserve trust and when is it perceived as illegitimate discretion? 
With approximately \$10 billion in technical exploit losses potentially addressable by onchain intervention (2016--2026), the design of these mechanisms has high practical stakes, but current approaches remain ad hoc and ideologically charged.
We address this gap by developing a \emph{Scope $\times$ Authority} taxonomy that maps the design space of emergency architectures along two dimensions: the precision of the intervention and the concentration of trigger authority.
We formalize the resulting tradeoffs of standing centralization cost, containment speed, and collateral disruption as a stochastic decision-support framework, and derive three empirical hypotheses from it.
Assessing the framework against 705 documented exploit incidents, we find that containment time varies systematically by authority type, that losses follow a heavy-tailed distribution ($\alpha \approx 1.33$) concentrating risk in rare catastrophic events, and that community sentiment plausibly modulates the effective cost of maintaining intervention capability. Using scope breadth as a practical proxy for blast potential, we also find that narrower interventions (Account/Module) do not underperform broader ones (Protocol/Network) on containment success and are slightly faster at the median, giving partial empirical support to the scope-blast hypothesis.
The analysis yields design guidance for emergency governance and reframes the problem as one of engineering tradeoffs rather than ideological debate.

\ifconcise
An extended version of this paper with detailed incident analysis, supplementary figures, and expanded discussion is available \href{https://github.com/e3o8o/legitimate-overrides-paper/blob/main/paper/main.pdf}{\textcolor{blue}{here}}.
\fi
\end{abstract}

\section{Introduction}\label{sec:intro}

Hacks and exploits are a persistent feature of blockchains and DeFi protocols, repeatedly producing losses that are large relative to protocol treasuries and TVL.
E.g., according to Charoenwong and Bernardi (2021; revised 2025)~\cite{CharoenwongBernardi2021Hacks}, cumulative losses from protocol failures, exploits, and market manipulation approaches \$88 billion. While much of this value derives from systemic market failures (e.g., Terra/Luna), a significant persistent strata ($\approx$ \$10 billion) consists of technical exploits potentially addressable by onchain emergency mechanisms.

A canonical early episode is the 2016 \emph{DAO} exploit~\cite{TheDAO2016}, whose aftermath culminated in a socially coordinated chain reconfiguration (the Ethereum hard fork)~\cite{EthereumDAOFork2016}, illustrating that ``immutability'' is ultimately mediated by governance when stakes are high.
In response, many systems embed \emph{emergency mechanisms} intended to limit damage under time pressure: protocol- or module-level pauses and shutdown procedures, issuer- or contract-level blacklisting/freeze controls, and (in some networks) chain-level transaction restrictions.
Such mechanisms can be effective in containment, but they introduce a legitimacy and centralization tension: intervention power creates an additional attack/abuse surface and changes the system's trust model; moreover, even when never exercised, the mere \emph{existence} of privileged override capability may reduce perceived trustlessness and thus depress utility or valuation.

\begin{remark}
Consistent with this, a recent large-scale scan of 166 blockchain networks reports that 16 chains contain active fund-freezing functions and another 19 could introduce similar capabilities with relatively minor changes (35/166 $\approx$ 21\%)~\cite{ByBit2025SecurityReport}; the same report highlights prominent deployments of these capabilities in practice, including the use of \emph{hardcoded} blacklists (addresses embedded directly in the chain's configuration, as used during the \$570M BNB Chain bridge exploit~\cite{BNBChainResponse}) and \emph{config-based} mechanisms (runtime-configurable lists, as deployed in the \$162M freeze of stolen assets on Sui after the Cetus hack~\cite{SuiCetusVote}).
\end{remark}

\paragraph{Immutability--intervention paradox.}
Emergency mechanisms are introduced to prevent catastrophic safety failures, yet they also create a second-order governance risk:
they alter the trust model by introducing privileged discretion (or privileged \emph{optionality}) over state transitions.
This yields an immutability--intervention paradox: in crises, communities often \emph{demand} intervention to protect users and integrated systems,
but outside crises, the same intervention capability can be viewed as a standing centralization backdoor whose very existence reduces credibility.

\paragraph{Safety--liveness tradeoff (SLT).}
We use the standard distributed-systems distinction: \emph{safety} prohibits ``bad'' state transitions (e.g., theft or invalid state),
whereas \emph{liveness} guarantees that ``good'' progress eventually happens (e.g., transactions continue to be processed).
Emergency overrides typically improve safety by restricting transitions, but thereby reduce liveness (and often censorship resistance).
Here, we are mainly interested in quantifying this tradeoff towards optimizing emergency mechanisms for different settings.

\paragraph{Contributions.}
These are our main contributions:
\begin{itemize}
    \item \textbf{Design-space taxonomy.} We propose a compact \emph{Scope $\times$ Authority} taxonomy that organizes emergency governance architectures
    into a single design space.
    \item \textbf{Incident mapping.} We document prominent override episodes (including chain-level responses and governance-led reconfigurations)
    and use them to ``fill'' the taxonomy, highlighting legitimacy tensions that recur across ecosystems.
    \item \textbf{Decision support framework.} We formalize the trade-offs between standing centralization and blast radius into a quantitative decision-support model and provide an open-source \emph{Intervention Mechanism Calculator}. 
    This tool enables protocol designers to calibrate mechanism selection based on community sentiment and estimated threat probabilities.\footnote{The tool is available \href{https://github.com/e3o8o/legitimate-overrides-paper/blob/main/notebooks/paper_analysis_complete.ipynb}{\textcolor{blue}{here}}.}
\end{itemize}

\ifconcise
\paragraph{Roadmap.}
\Cref{sec:related} positions our work within relevant literatures.
\Cref{sec:taxonomy} presents our Scope $\times$ Authority taxonomy.
\Cref{sec:theory} formalizes the decision problem as a stochastic optimization.
\Cref{sec:empirical} assesses the model with comprehensive empirical analysis.
\Cref{sec:implications} extracts design principles for practitioners.
\else
\paragraph{Roadmap.}
\Cref{sec:related} positions our work within relevant literatures.
\Cref{sec:democracy-analogy} draws on constitutional democratic theory for conceptual foundations.
\Cref{sec:taxonomy} presents our Scope $\times$ Authority taxonomy.
\Cref{sec:theory} formalizes the decision problem as a stochastic optimization.
\Cref{sec:empirical} assesses the model with comprehensive empirical analysis.
\Cref{sec:implications} extracts design principles for practitioners.
\fi

\section{Related Work}
\label{sec:related}

Our work intersects four bodies of literature.
\paragraph{Blockchain security and incident response.}
The security community has documented exploit patterns and proposed defensive mechanisms~\cite{CharoenwongBernardi2021Hacks}. Recent comprehensive reviews by Dwivedi et al.~\cite{Dwivedi2024BlockchainAttacks} and Siam et al.~\cite{Siam2025SecuringBlockchain} systematically categorize vulnerabilities across blockchain layers -- from network-level attacks to smart contract exploits -- yet most work focuses on documentation and \emph{ex-post} analyses rather than prevention, preemption, or post-incident governance. Notable exceptions include analyses of the DAO fork~\cite{ETCAnnouncement2016} and studies of bug bounty programs as coordinated disclosure mechanisms~\cite{Badash2021BugBounty}.
Importantly, \emph{legitimate overrides} and the broader design of formal intervention mechanisms remain significantly under-studied in both academic and practitioner literature; our work aims to fill this gap by providing a quantitative framework.

\paragraph{Pre-execution prevention primitives.}
\label{subsec:pre-execution-prevention-primitives}
An emergent class of mechanisms operates \emph{before} transaction execution, shifting the intervention point from reactive pauses to deterministic pre-execution enforcement. The Phylax Credible Layer~\cite{PhylaxCredibleLayer2026} exemplifies this paradigm: protocols define ``assertions'' -- Solidity-written invariants that specify states which should never occur (e.g., ``a healthy account cannot become liquidatable in a single transaction''). The network's sequencer then simulates every transaction against registered assertions and drops violations before execution. This architecture inherits the chain's trust assumptions (no external oracle or monitoring service) and provides zero false positives by construction. Early adopters include Euler Finance (``Holy Grail'' invariant for account liquidity)~\cite{PhylaxEulerAssertions2026}, Malda Protocol, Turtle/Lagoon, and Denaria, each deploying 1--6 assertions to protect critical protocol invariants. 
Linea's integration of the Credible Layer into its sequencer (January 2026)~\cite{PhylaxCredibleLayer2026} demonstrates institutional appetite for infrastructure-level security guaranties. 
This primitive occupies a distinct cell in our taxonomy: Network Scope (sequencer-level enforcement) with Signer Set Authority (sequencer operator controls assertion enforcement), offering a compelling alternative to reactive pauses in risk-averse environments. Unlike reactive mechanisms that respond after exploit detection, pre-execution prevention operates deterministically at the transaction level before execution, representing a fundamentally different paradigm that extends beyond our core 5$\times$3 framework of reactive interventions (which operate at network, asset, protocol, module, or account scope).

\paragraph{DAO governance.}
Recent empirical work examines onchain governance mechanisms, voting behavior, and the tension between efficiency and decentralization~\cite{Wang2025DAOempirical,Ma2025DAOsecurity,Werbach2024BlockchainGovernance}. Wang et al.~\cite{Wang2025DAOempirical} analyzed 581 DAOs and 16,246 proposals to document governance dynamics, while Ma et al.~\cite{Ma2025DAOsecurity} examined 3,348 DAOs across 9 blockchains, revealing widespread governance vulnerabilities including contract backdoors and malicious proposals. Qian~\cite{Qian2025DAOgovernance} demonstrated how flash loan attacks, offchain voting manipulation, and token-based coercion led to over \$300M in losses across major DAOs. Werbach et al.~\cite{Werbach2024BlockchainGovernance} surveyed 23 blockchain projects to compare onchain and offchain governance practices, highlighting the gap between pristine abstractions and messy realities. Our focus on \emph{emergency} governance-decisions under time pressure complements this literature.

\paragraph{Political science of emergency powers.}
The political science literature on states of emergency, constitutional constraints, and executive overreach provides conceptual tools for analyzing blockchain emergency mechanisms. Scholars have long studied how democracies balance speed and accountability during crises, a tension that maps directly to our Authority dimension~\cite{ferejohn2004law, gimpelson2016law}. We bridge this connection to open a dialogue between blockchain governance and formal constitutional theory of exception.

\ifconcise\else
\section{Relation to Constitutional Democracies}
\label{sec:democracy-analogy}

Emergency override capabilities in blockchains are often discussed as technical ``circuit breakers'' or
governance exceptions. A closely related debate exists in constitutional democracies, where emergency
powers are introduced to address rare, high-stakes crises while preserving the legitimacy of the
ordinary constitutional order. The parallel is substantive: in both settings, emergencies intensify the
tradeoff between \emph{speed and containment} on the one hand, and \emph{constraint and legitimacy} on
the other. Our goal is not a full comparative political-theory analysis, but rather to (i) acknowledge
this established body of work, and (ii) use it as conceptual inspiration for organizing and analyzing
blockchain emergency mechanisms.

\paragraph{A minimal taxonomy in constitutional democracies.}
A useful starting point is a two-part taxonomy, common in the law-and-politics literature on
emergencies:
\begin{enumerate}
    \item \textbf{Emergency legislation (ad hoc statutes).}
    The legislature enacts special, typically time-limited rules aimed at a specific crisis (e.g.,
    public-health emergencies such as COVID-19). This approach keeps the ordinary constitutional structure formally
    ``on,'' while temporarily expanding executive capabilities via statute.
    \item \textbf{Declared exceptional regime (state of emergency).}
    A formally declared exceptional state activates extraordinary powers that would be unavailable
    under normal conditions. A canonical normative rationale is \emph{conservative/commissarial}:
    extraordinary powers are justified to neutralize the threat and restore the ordinary order ``as
    soon as possible,'' rather than to permanently reconstitute the regime
   ~\cite{ferejohn2004law} (the canonical example would be Rome's dictator).
\end{enumerate}
A recurring conceptual issue is \emph{declaration authority}: whether the exception is treated as
objectively detectable (threshold-triggered) or as requiring an institution with epistemic authority to
declare that an exceptional situation exists~\cite{ferejohn2004law}.

\paragraph{Canonical risks and safeguards.}
Even when emergency powers are formally enacted, the democratic literature emphasizes that emergency
governance is not ``lawless''; rather, it is designed (or at least justified) through constraints and
procedures. Two themes are particularly relevant for our setting.

\begin{itemize}
    \item \textbf{Stickiness (ratchet risk).}
    A central concern is that extraordinary powers, once activated, may persist through repeated
    renewals or institutional drift. Comparative evidence illustrates how formally temporary
    declarations can become routine; for example, Israel's general state of emergency has been
    repeatedly renewed since 1948, and the continued validity of various legal provisions may depend
    on the declaration, creating structural incentives for renewal~\cite{gimpelson2016law}.
    \item \textbf{Boundedness principles (legality, proportionality, purpose limitation).}
    Many constitutional systems articulate principles limiting emergency action, including
    exceptionality/last-resort, legality (no implied powers), proportionality, and purpose limitation
    (actions should aim at combating the threat and restoring ordinary functioning). For instance, a
    synopsis of the Polish framework lists these principles explicitly~\cite{gimpelson2016law}.
    Similarly, Israel's emergency regulations regime contains explicit restrictions tied to
    proportionality/necessity (``only to the extent warranted by the state of emergency'') and
    protections for core rights~\cite{gimpelson2016law}.
\end{itemize}
\fi

\section{Taxonomy: Scope \texorpdfstring{$\times$}{x} Authority}
\label{sec:taxonomy}

Our central organizing device is a two-dimensional taxonomy of emergency mechanisms along two orthogonal design dimensions:
(i) \emph{Scope} (the precision / blast radius of the intervention), and (ii) \emph{Authority} (who can trigger it, and how).
The safety--liveness profile is treated as an induced property of these design choices, rather than an axis in itself.

\subsection{Dimension I: Scope (Hierarchy of Precision)}
\label{sec:taxonomy-scope}

We model scope as a discrete hierarchy of precision levels. Moving downward increases precision (reduces blast radius)
and typically reduces collateral disruption, but may require stronger instrumentation and may increase response complexity.

\begin{enumerate}
    \item \textbf{Network scope.} Chain-wide restriction or reconfiguration affecting all applications (e.g., halt/pause, chain-wide censorship rules, reorg/rollback, global fork-based remediation).
    \item \textbf{Asset scope.} Actions targeting a specific asset across holders/venues (e.g., issuer blacklisting, asset-specific freezes/burns, bridge-wide caps on a given token).
    \item \textbf{Protocol scope.} Application-wide restriction within a specific protocol (e.g., pausing all markets in a lending protocol; emergency shutdown of a stablecoin system).
    \item \textbf{Module scope.} Feature-specific restriction within a protocol (e.g., pausing liquidations while allowing deposits/repayments; pausing oracle updates).
    \item \textbf{Account scope.} Targeted restriction or remediation affecting specific addresses/accounts only (e.g., freezing or quarantining addresses implicated by evidence).
\end{enumerate}

\subsection{Dimension II: Authority (Trigger Holder)}
\label{sec:taxonomy-authority}

We distinguish three authority modes, ordered from concentrated to broadly distributed:

\begin{enumerate}
    \item \textbf{Signer set (key-based).} A fixed keyholder set (e.g., 1-of-$n$ or $m$-of-$n$ multisig) can trigger the mechanism.
    \item \textbf{Delegated body.} A designated council/committee (typically multi-party) holds bounded emergency powers, often with mandates, reporting duties, and ex post accountability.
    \item \textbf{Governance process.} The intervention requires a formal vote or a broadly coordinated social process (e.g., token-holder governance, validator/community coordination for upgrades).
\end{enumerate}

\ifconcise\else
\begin{remark}[Political Analogy]
This authority spectrum maps naturally to classical political theory classifications: \textbf{Signer Set} corresponds to \emph{Oligarchy} (power vested in a small, self-selected or appointed group); \textbf{Delegated Body} corresponds to \emph{Representative Democracy} (power exercised by elected representatives within constitutional bounds); and \textbf{Governance Process} corresponds to \emph{Direct Democracy} (power exercised directly by the citizenry/token-holders). We use the technical nomenclature throughout the main text for precision, but the political analogy illuminates the legitimacy trade-offs inherent in each mode.
\end{remark}
\fi

\begin{table}[!htbp]
\centering
\scriptsize
\setlength{\tabcolsep}{3pt}
\renewcommand{\arraystretch}{1.0}

\begin{tabularx}{\linewidth}{
  >{\raggedright\arraybackslash}p{2.0cm}
  >{\raggedright\arraybackslash}X
  >{\raggedright\arraybackslash}X
  >{\raggedright\arraybackslash}X
}
\toprule
\textbf{Scope $\backslash$ Authority} &
\textbf{Signer set} &
\textbf{Delegated body} &
\textbf{Governance process} \\
\midrule

\textbf{Network}
&
\emph{Key-triggered chain-wide restriction} \newline
(e.g., halt/pause; global censorship toggle) \newline
\textbf{Examples:} Harmony Horizon (Jun 2022, \$100M)~\cite{Harmony2022Exploit}; BNB Chain halt (Oct 2022, \$570M)~\cite{ReutersBNB2022}; Berachain halt (Nov 2025)~\cite{Berachain2025PostMortem}
&
\emph{Council-coordinated network action} \newline
(e.g., bounded emergency council mandate) \newline
\textbf{Examples:} Poly Network validator coordination (Aug 2021, \$611M returned)~\cite{PolyNetworkElliptic}
&
\emph{Governance-led chain reconfiguration} \newline
(e.g., coordinated upgrade/fork-based remediation) \newline
\textbf{Examples:} Ethereum DAO fork (Jun 2016)~\cite{TheDAO2016}; Gnosis Chain hard fork (Dec 2025, \$9.4M)~\cite{Gnosis2025HardFork} \\
\midrule

\textbf{Asset}
&
\emph{Issuer/admin asset controls} \newline
(e.g., blacklist/freeze/burn hooks) \newline
\textbf{Examples:} Tether USDT freezes (PDVSA \$182M)~\cite{TetherVenezuela2026}; Circle USDC blocked addresses (Tornado Cash \$75K)~\cite{CircleUSDCTerms2025}; WLFI blacklist of Justin Sun (\$107M)~\cite{WLFIJustinSun2025}
&
\emph{Delegated asset committee} \newline
(e.g., bridge/operator council enforces caps/blocks) \newline
\textbf{Examples:} Gnosis Bridge Board freeze (Nov 2025)~\cite{Gnosis2025BalancerHack}; Curve Emergency DAO (emissions/PSR only)~\cite{CurveExploitLlamaRisk}
&
\emph{Governance changes asset rules} \newline
(e.g., parameter change, migration, social recovery) \newline
\textbf{Examples:} Yearn governance-approved pxETH burn~\cite{Yearn2025Recovery} \\
\midrule

\textbf{Protocol}
&
\emph{Admin pause / shutdown} \newline
(e.g., protocol-wide circuit breaker) \newline
\textbf{Examples:} Balancer CSPv6 auto-pause (Nov 2025)~\cite{BalancerPostMortem2025}; Liqwid Kill Switch~\cite{Liqwid2024Governance}; Beanstalk shutdown (Apr 2022, \$182M); Superfluid agreement halt (Feb 2022)
&
\emph{Security-council pause} \newline
(e.g., bounded emergency mandate) \newline
\textbf{Examples:} Aave Protocol Guardians~\cite{AaveGuardians}; Balancer V3 Emergency subDAO~\cite{BalancerV3EmergencySubDAO}; Liqwid Pause Guardian (4-of-X)
&
\emph{DAO-administered emergency action} \newline
(e.g., vote to pause/upgrade/settle) \newline
\textbf{Examples:} MakerDAO ESM (deprecated)~\cite{MakerDAOEmergencyShutdown}; Anchor Protocol (defunct)~\cite{Anchor2022} \\
\midrule

\textbf{Module}
&
\emph{Admin disables a feature} \newline
(e.g., stop liquidations / withdrawals) \newline
\textbf{Examples:} Compound price oracle pauses (2021); FEI/Rari DAI borrow pause (Apr 2022, \$80M); Elephant Money TRUNK minting pause (Apr 2022)
&
\emph{Delegated feature-specific pauses} \newline
(e.g., guardian pauses liquidations) \newline
\textbf{Examples:} Aave V3 reserve/pool pauses~\cite{AaveACLManager}; Liqwid market pause (Oct 2025); dYdX YFI circuit breaker~\cite{dYdXSushiYFI}
&
\emph{Governance toggles module parameters} \newline
(e.g., vote to enable / disable a feature temporarily) \newline
\textbf{Examples:} MakerDAO (Sky) USDC-PSM pause (Mar 2023)~\cite{MakerDAOEmergency2023}; Aave asset parameter updates~\cite{AaveDPIFreezeProposal}; Solend USDH LTV set to 0 (Nov 2022) \\
\midrule

\textbf{Account}
&
\emph{Key-based targeted restriction} \newline
(e.g., freeze/quarantine addresses) \newline
\textbf{Examples:} Tether/Circle address blacklists; Sonic freezeAccount (Nov 2025)~\cite{Sonic2025Freeze}; Ronin attacker freeze~\cite{Ronin2022Exploit}
&
\emph{Delegated targeted remediation} \newline
(e.g., council-authorized quarantines) \newline
\textbf{Examples:} StakeWise multisig burn/mint (\$20.7M)~\cite{StakeWise2025Recovery}; Flow Isolated Recovery (Dec 2025, 1,060 addresses)~\cite{Flow2025Recovery}
&
\emph{Governance-authorized targeted action} \newline
(e.g., vote-based remediation against identified addresses) \newline
\textbf{Examples:} Sui/Cetus 90.9\% stake vote (May 2025, \$162M)~\cite{SuiCetusVote}; VeChain blocklist (Dec 2019)~\cite{VeChain2025Refutation} \\
\bottomrule
\end{tabularx}

\caption{\textbf{Emergency mechanisms mapped to Scope (precision) $\times$ Authority (trigger holder).} The table defines the design space used to structure the narrative evidence and, later, the formal analysis. Some incidents span cells (e.g., Sui/Cetus: Delegated Body freeze $\to$ Governance recovery vote).}
\label{tab:scope-authority}
\end{table}

\ifconcise\else
\subsection{Intuition for the Taxonomy}
\label{sec:taxonomy-why}

Two basic intuitions motivate the taxonomy.
First, \emph{scope} governs collateral disruption: higher-precision interventions affect fewer uninvolved users and contracts,
but typically require stronger instrumentation and more careful evidence.
Second, \emph{authority} governs speed and contestability: concentrated authority can respond quickly 
but increases perceived discretion; governance-heavy authority can increase legitimacy and accountability 
but may be too slow for rapidly unfolding incidents.

\Cref{tab:scope-authority} ``fills'' the design space with prominent episodes and isolates recurring legitimacy tensions that appear across ecosystems and across cells of the table. The empirical analysis in \Cref{sec:empirical} examines these cases in detail.

\subsubsection{Relation to Constitutional Democracies}

We discuss a bit more our taxonomy and contribution with relation to constitutional democracies.

\paragraph{Mapping to the blockchain setting.}
These democratic themes naturally translate to decentralized protocols:
\begin{itemize}
    \item \textbf{Authority $\leftrightarrow$ who may trigger the exception.}
    Democratic debates about the locus of emergency authority (executive discretion vs. legislative
    delegation and oversight) correspond to our \emph{Authority} dimension: concentrated keyholder
    triggers, delegated emergency bodies, and full governance processes.
    \item \textbf{Scope $\leftrightarrow$ proportionality and collateral impact.}
    The constitutional demand for proportionality/purpose limitation corresponds to our \emph{Scope}
    dimension: more precise mechanisms reduce blast radius and collateral disruption, but may require
    stronger instrumentation and evidence.
\end{itemize}
In this sense, our Scope $\times$ Authority taxonomy can be viewed as a protocol-design analogue of
core questions studied in constitutional emergency governance: \emph{who is empowered to act}, and
\emph{how far the intervention may reach}.

\paragraph{Positioning our contribution.}
In the blockchain space, emergency mechanisms are widespread but heterogeneous (pauses, freezes,
emergency councils, forks/rollbacks), and design choices are often justified ad hoc, incident by
incident. Here we impose structure by (i) organizing mechanisms into a unified design space
(Scope $\times$ Authority), and (ii) connecting design choices to explicit tradeoffs.
\fi

\section{A Stochastic Model of Emergency Governance}
\label{sec:theory}

The Scope $\times$ Authority taxonomy (\Cref{sec:taxonomy}) suggests that emergency governance is, fundamentally,
a \emph{design problem under time pressure and uncertainty}.
A protocol must choose ex ante which override capabilities to embed (if any), how precise they can be, and who can trigger them;
then, when an incident occurs, the chosen architecture constrains which responses are feasible and how quickly they can be executed.
The empirical cases in \Cref{subsec:detailed-cases} demonstrate that similar threat events yield very different interventions, and that perceived legitimacy depends not only on outcomes but also on scope, authority, and procedural safeguards.

\paragraph{Design objective (informal).}
At a high level, an emergency mechanism trades off three ingredients:
(i) \emph{containment} of losses from an unfolding incident (which typically favors fast, powerful triggers),
(ii) \emph{collateral disruption} to uninvolved users and applications (which typically favors higher precision),
(iii) the \emph{standing centralization cost} of privileged override capability (which is incurred even when no emergency occurs,
since it changes the trust model and increases perceived discretion).

\subsection{A Minimal Stochastic Model}

\newcommand{\Events}{\mathcal{H}}          
\newcommand{\Prob}{\Pr}                    
\newcommand{\Time}{\mathrm{Time}}          
\newcommand{\DamageRate}{\mathrm{DamageRate}} 
\newcommand{\BlastRate}{\mathrm{BlastRate}}   
\newcommand{\CentralizationCost}{\mathrm{CentralizationCost}} \newcommand{\ExpectedCost}{\mathrm{ExpectedCost}}

A protocol designer chooses an emergency governance architecture $m\in\mathcal{M}$.
Future adverse events are modeled by a finite set of types $\Events$ and a distribution $\Prob[\cdot]$ over $\Events$
(optionally including a ``no-incident'' type; think of it as a static decision for the next timestep).

If an event of type $h\in\Events$ occurs and is not yet contained, it generates loss at rate $\DamageRate(h)\ge 0$ per unit time (the ``no-incident'' type has $\DamageRate(h) = 0$). Architecture $m$ contains the event after $\Time(m)\ge 0$ time units. Exercising $m$ also induces a collateral disruption cost $\BlastRate(m)\ge 0$ (capturing the fixed, one-time shock of its scope). Finally, $m$ imposes a standing centralization cost $\CentralizationCost(m)\ge 0$, incurred regardless of whether an incident occurs (Chekhov's gun).

Thus, given a distribution over bad-event types $h\in\Events$ -- specified by their probabilities $\Prob[h]$ and damage rates $\DamageRate(h)$ -- and given an emergency governance architecture $m\in\mathcal{M}$ with containment time $\Time(m)$, standing centralization cost $\CentralizationCost(m)$, and blast-radius cost $\BlastRate(m)$, the designer's task is to minimize the expected cost.
We define the expected cost of architecture $m$ by
\begin{align*}
\ExpectedCost(m)
&\;:=\;
\CentralizationCost(m) \\
&\;+\;
\sum_{h\in\Events}\Prob[h]\cdot
\big(\Time(m)\cdot\DamageRate(h) + \BlastRate(m)\big),
\end{align*}
and study the design problem
\[
\min_{m\in\mathcal{M}} \ExpectedCost(m).
\]

\ifconcise\else
\paragraph{Interpretation.}
The blast rate $\BlastRate(m)$ is a \emph{one-time} cost because the scope decision is about width, not duration: pausing a protocol or chain freezes activity causing a one-time shock: indeed, extending a pause from, e.g., 1 hour to 3 hours does not make the blast radius 3$\times$ larger. If we multiplied Blast Rate by Time continuously, a Governance decision taking 3 days versus a Signer Set decision in 30 minutes would translate mathematically as 144$\times$ more destructive, even though the impact of intervention is effectively the same as a one-time event.
\fi

\subsection{Three Theoretical Predictions}
\label{subsec:predictions}

Our model yields three testable predictions that we assess in \Cref{sec:empirical}:

\begin{enumerate}
    \item \textbf{Prediction 1 (Speed-Centralization Tradeoff):} Faster architectures (Signer Set) minimize exploit losses but impose higher standing centralization costs than slower architectures (Governance).
    \item \textbf{Prediction 2 (Scope-Blast Relationship):} Higher-precision interventions (Account, Module scope) achieve comparable containment outcomes with lower practical blast potential than broader interventions (Protocol, Network scope). In the empirical section, we operationalize blast potential through scope breadth as a proxy, rather than claiming a direct market-wide collateral-loss measure.
    \item \textbf{Prediction 3 (Sentiment-Cost Modulation):} Positive community sentiment toward emergency mechanisms reduces their effective standing centralization cost; negative sentiment increases it.
\end{enumerate}

\section{Empirical Analysis and Assessment}
\label{sec:empirical}

Building on the theoretical framework in \Cref{sec:theory}, we now present comprehensive empirical evidence using data from 705 documented exploit incidents (2016--2026). Motivated by Pearson's Law that \emph{``that which is measured and reported improves exponentially,''} we develop measurement frameworks to illuminate how legitimate emergency interventions can reduce protocol losses and protect users.

\ifconcise\else
\begin{remark}[Datasets]
We draw on multiple complementary datasets. First, we aggregate data on 705 documented exploit incidents from 2016--2026, recording date, chain, loss magnitude, and attack vector for each event. From these 705 total cases, we identify 640 technical exploits. Of these technical cases, 601 represent our intervention-eligible universe -- technical exploits where emergency mechanisms could theoretically apply. The remaining 39 technical cases lack viable intervention points and are excluded from effectiveness analysis. From these 601 intervention-eligible cases, 130 involve actual emergency mechanism activations (reactive responses to exploits and proactive measures), representing the complete universe of intervention responses in our dataset.

Second, we use a curated \emph{Intervention Incidents} subset of 52 high-fidelity cases with verified timing, authority type, scope, and outcome data. This subset -- which includes both reactive exploit responses and proactive interventions -- represents the highest-quality cases selected for detailed effectiveness analysis. This subset was constructed through manual verification of incident reports, cross-referencing with multiple security databases, and validation of intervention details through official post-mortems and governance forum discussions.
\end{remark}
\fi

\subsection{Dataset Overview and Stratification}
\label{subsec:stratification}

A critical distinction in our analysis is the separation between ``systemic failures'' and ``intervention-eligible exploits.'' We stratify the 705 documented cases into four categories:

\begin{enumerate}
    \item \textbf{Systemic Failures} (10 cases, \$61.80B): Massive economic design collapses (e.g., Terra/Luna, FTX) where no emergency pause mechanism could prevent loss.
    \item \textbf{Other Non-Addressable} (94 cases, \$7.41B): Incidents like rug pulls, phishing, or unpausable logic bugs where intervention was not technically feasible.
    \item \textbf{Intervention-Eligible} (601 cases, \$9.60B): Technical exploits (reentrancy, logic bugs, etc.) where emergency mechanisms were applicable.
    \item \textbf{Actually Intervened} (130 cases, \$7.51B): Cases where emergency mechanisms were actually activated.
\end{enumerate}

Our effectiveness analysis focuses strictly on the \textbf{601 intervention-eligible cases}. Including systemic failures (as done in raw aggregators) would severely distort the analysis, as a \$40B collapse like Terra is not ``addressable'' by the emergency governance mechanisms we study.

\ifconcise\else
\begin{figure}[t]
    \centering
    \includegraphics[width=0.9\linewidth]{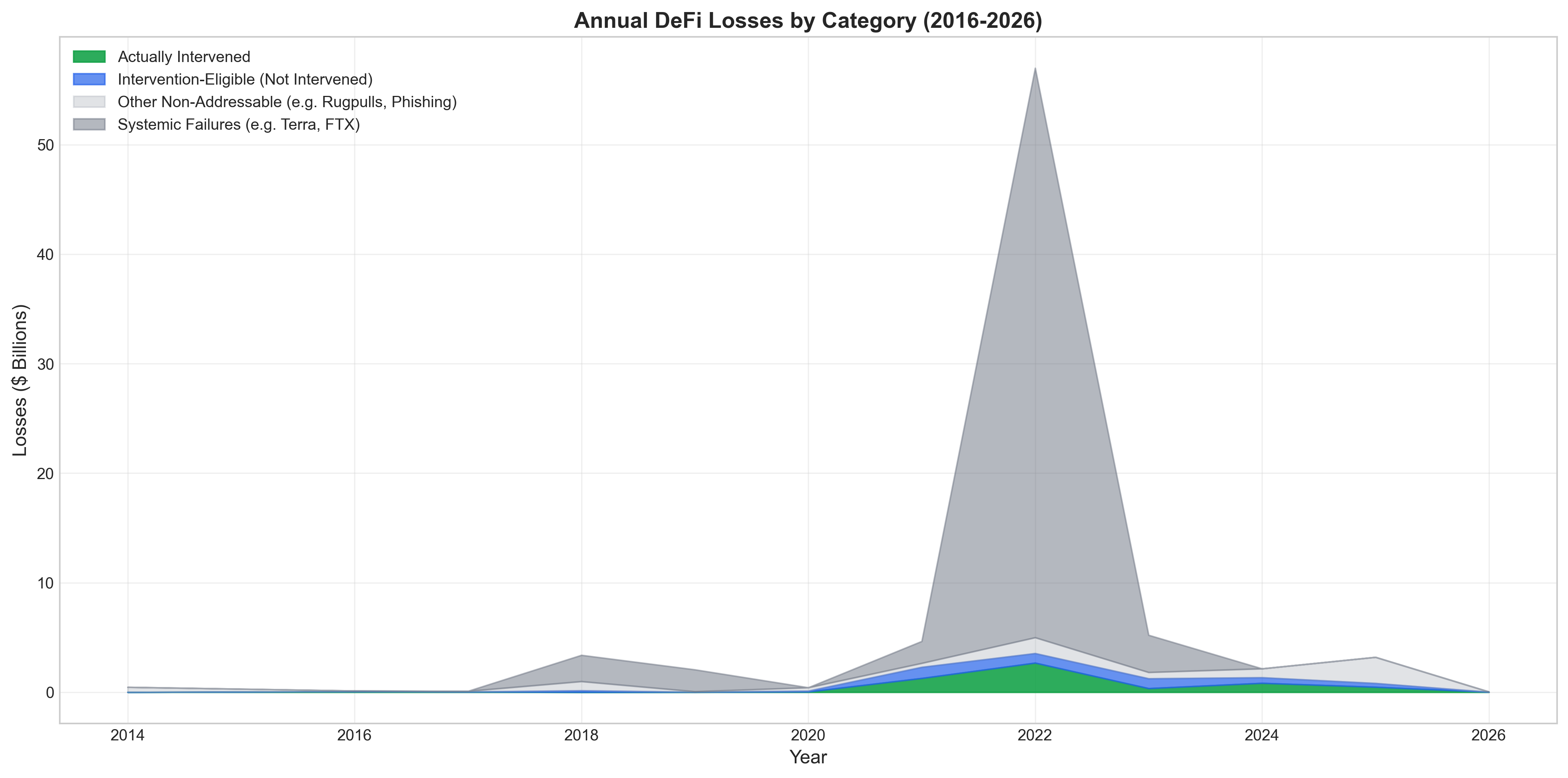}
    \caption{\textbf{Stratification of Losses (2016-2026).} We stratify losses into four layers: \textbf{Systemic Failures} (dark grey, e.g., Terra), \textbf{Other Non-Addressable} (light grey, e.g., rug pulls), \textbf{Intervention-Eligible} (blue), and \textbf{Actually Intervened} (green). This reveals that while systemic events dominate 2022, addressable technical exploits represent a consistent baseline of risk.}
    \label{fig:timeline_stratification}
\end{figure}
\fi

\subsection{Loss Distribution: Power Law Validation}
\label{subsec:power-law}

The resulting distribution of losses (\Cref{fig:pareto_loss}) follows a power law (Kolmogorov-Smirnov test statistic $D=0.150$, $p < 0.001$, Power Law exponent $\alpha \approx 1.33$), confirming that risk is driven by fat-tail events. This Pareto-like pattern (approximately 80\% of cumulative losses from fewer than 50 incidents) implies that the expected value of intervention capability is driven primarily by its effectiveness against ``super-hacks,'' where rapid containment can prevent tens or hundreds of millions in additional losses.

\ifconcise\else
\begin{figure}[t]
    \centering
    \includegraphics[width=0.85\linewidth]{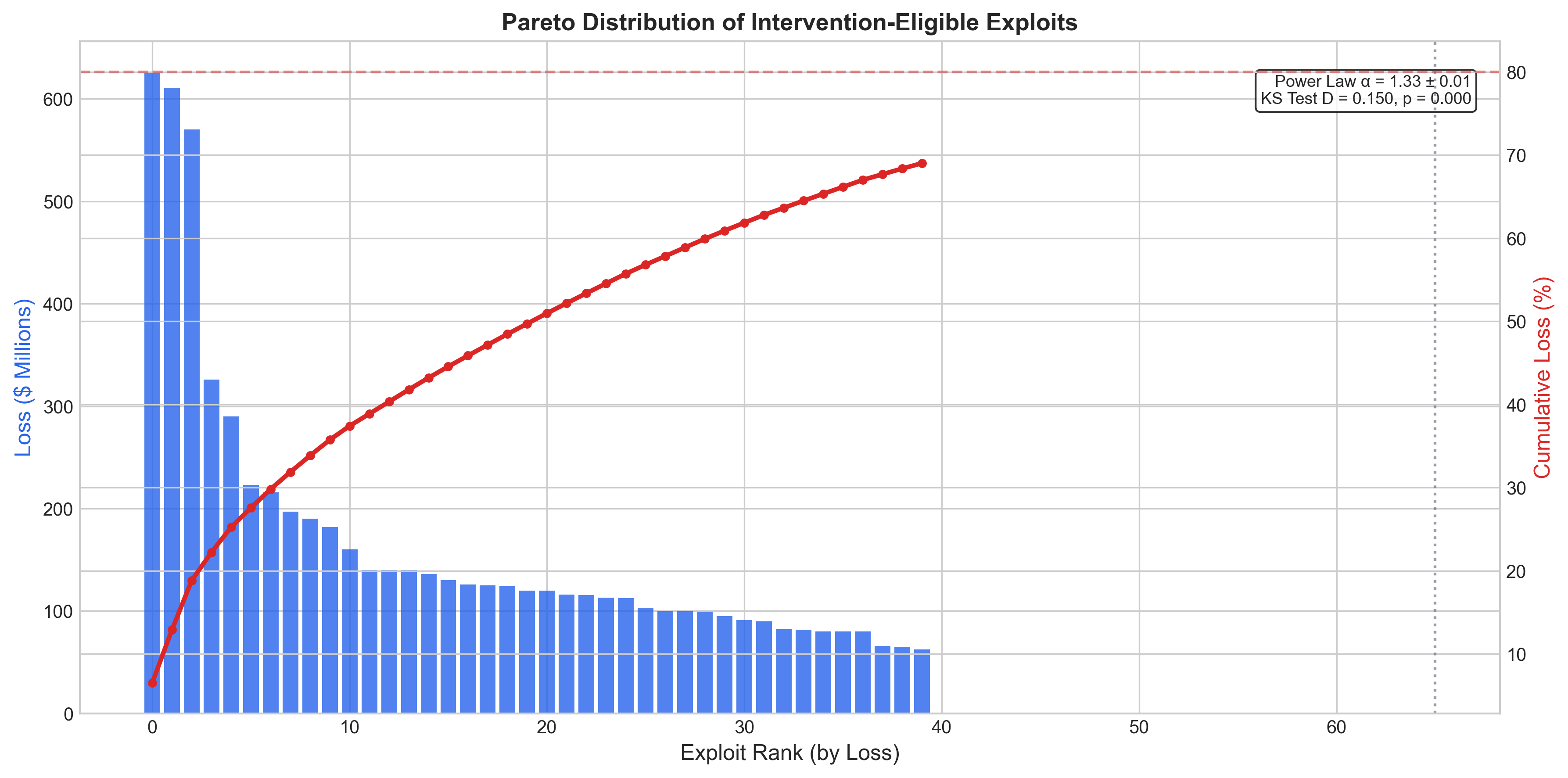}
    \caption{\textbf{Pareto Distribution of Intervention-Eligible Losses.} Approximately 80\% of cumulative losses in our addressable dataset are attributable to fewer than 50 incidents. This extreme concentration implies that intervention capability is most valuable against rare, catastrophic events (``super-hacks''). Note that this chart excludes \$70B+ in systemic economic failures (e.g., Terra, FTX) which are not addressable by emergency overrides. Power law fit: $\alpha \approx 1.33$, KS test $D=0.150$, $p < 0.001$.}
    \label{fig:pareto_loss}
\end{figure}
\fi

\ifconcise\else
\Cref{fig:top_10_exploits} shows the breakdown of the largest technical exploits, revealing that a handful of ``super-hacks'' drive the vast majority of preventable losses, reinforcing the power law finding. Stacked bars show losses prevented (green) versus lost (red).

\begin{figure}[t]
    \centering
    \includegraphics[width=0.85\linewidth]{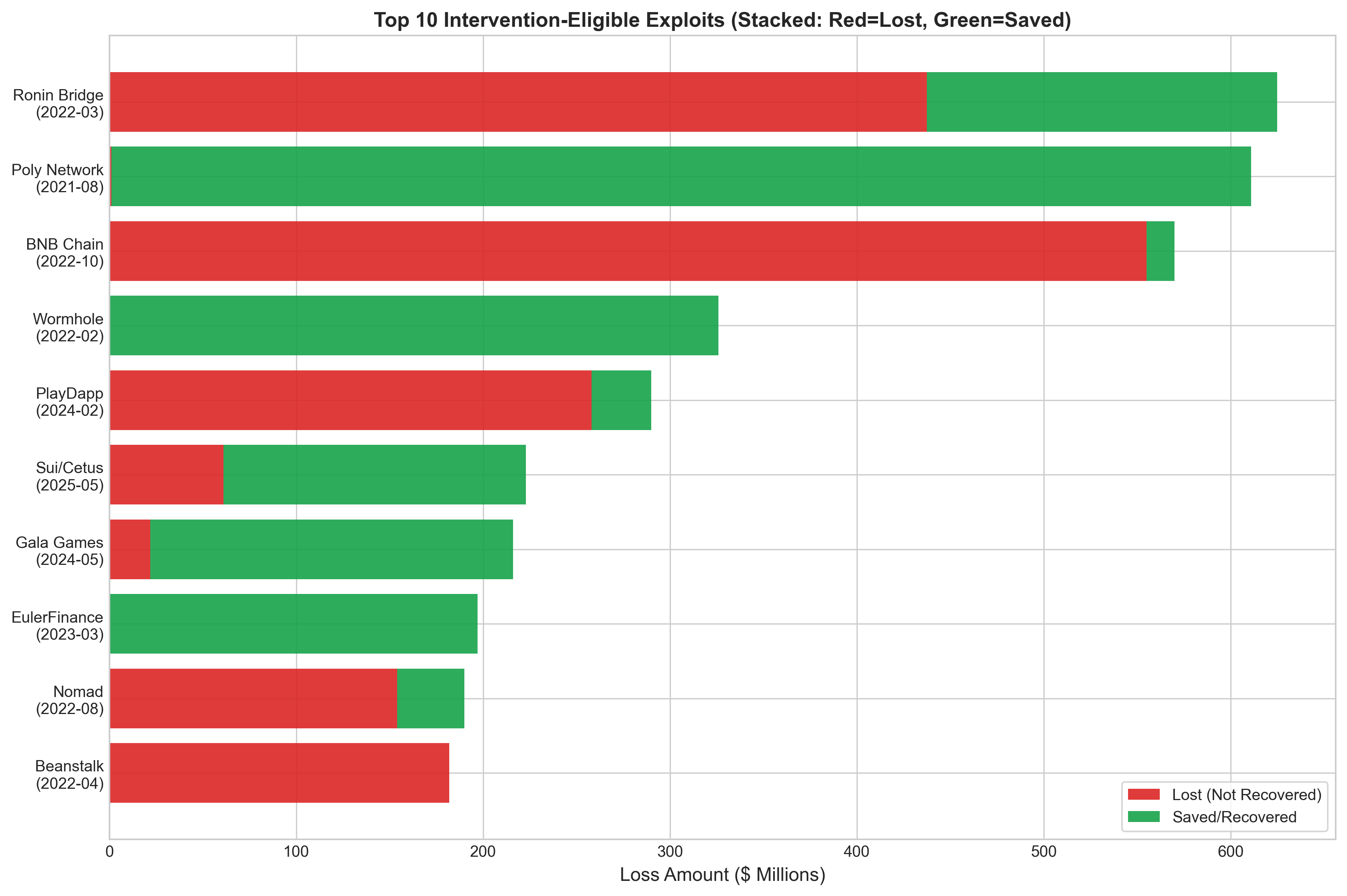}
    \caption{\textbf{Top 10 Intervention-Eligible Exploits.} The breakdown of the largest technical exploits reveals that a handful of ``super-hacks'' drive the vast majority of preventable losses, reinforcing the power law finding. Stacked bars show losses prevented (green) versus lost (red).}
    \label{fig:top_10_exploits}
\end{figure}
\fi

\ifconcise\else
\subsection{Attack Vector Characterization}
\label{subsec:attack-vectors}

\begin{figure}[t]
    \centering
    \includegraphics[width=0.85\linewidth]{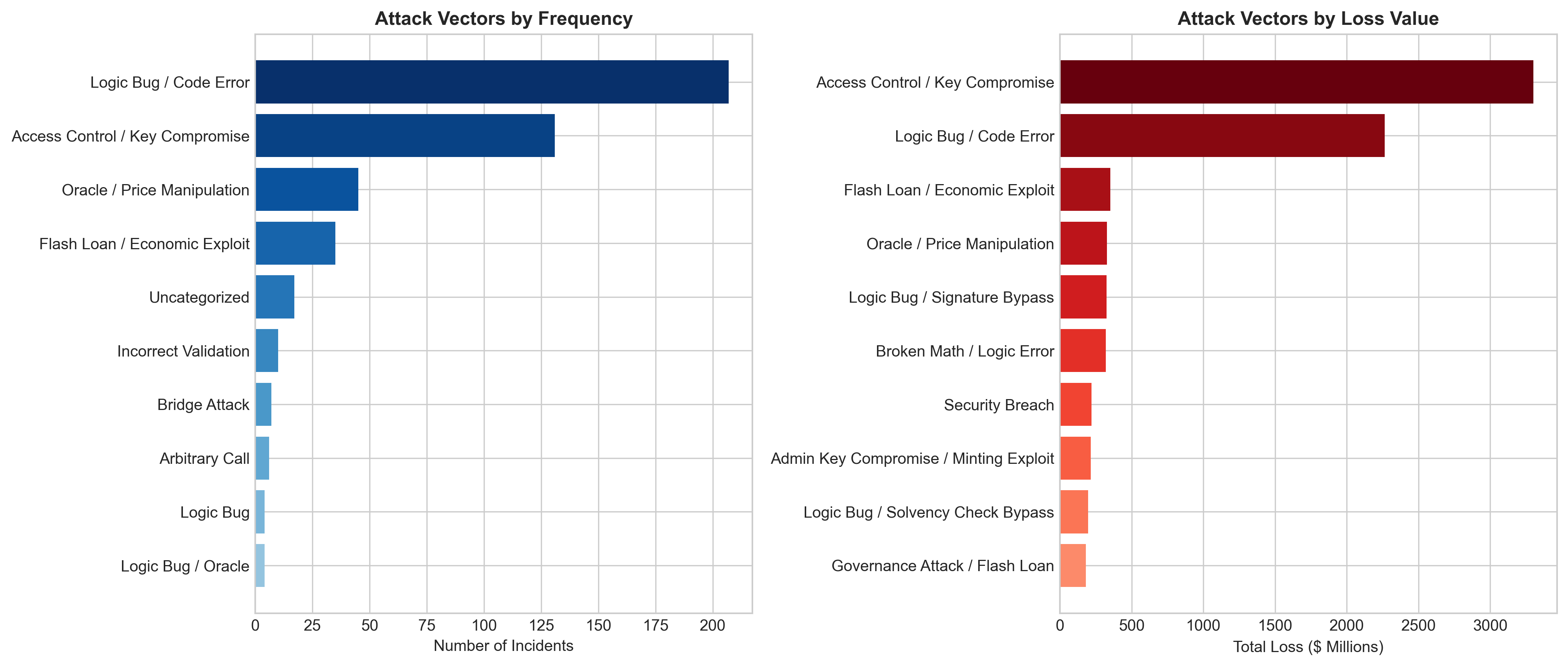}
    \caption{\textbf{Attack Vector Distribution.} We observe that while `Logic Errors' and `Access Control' issues are frequent and account for significant losses; complex `Oracle Manipulation' and `Flash Loan' attacks often result in the highest severity incidents, necessitating rapid intervention capabilities.}
    \label{fig:attack_vectors}
\end{figure}

\Cref{fig:attack_vectors} shows the distribution of attack vectors by frequency (left panel) and total loss value (right panel). While Logic Errors and Access Control issues are most frequent and cause significant damage, Oracle Manipulation and Flash Loan attacks come next among the highest severity incidents. This has direct implications for mechanism design: protocols vulnerable to flash loan attacks (which unfold in a single transaction block) require fastest-response mechanisms (Signer Set or Delegated Body), while slower Governance processes may suffice for vulnerability types with longer exploitation windows.
\fi

\subsection{The Speed-Centralization Tradeoff}
\label{subsec:authority-performance}

\ifconcise\else
\begin{figure}[t]
    \centering
    \includegraphics[width=0.85\linewidth]{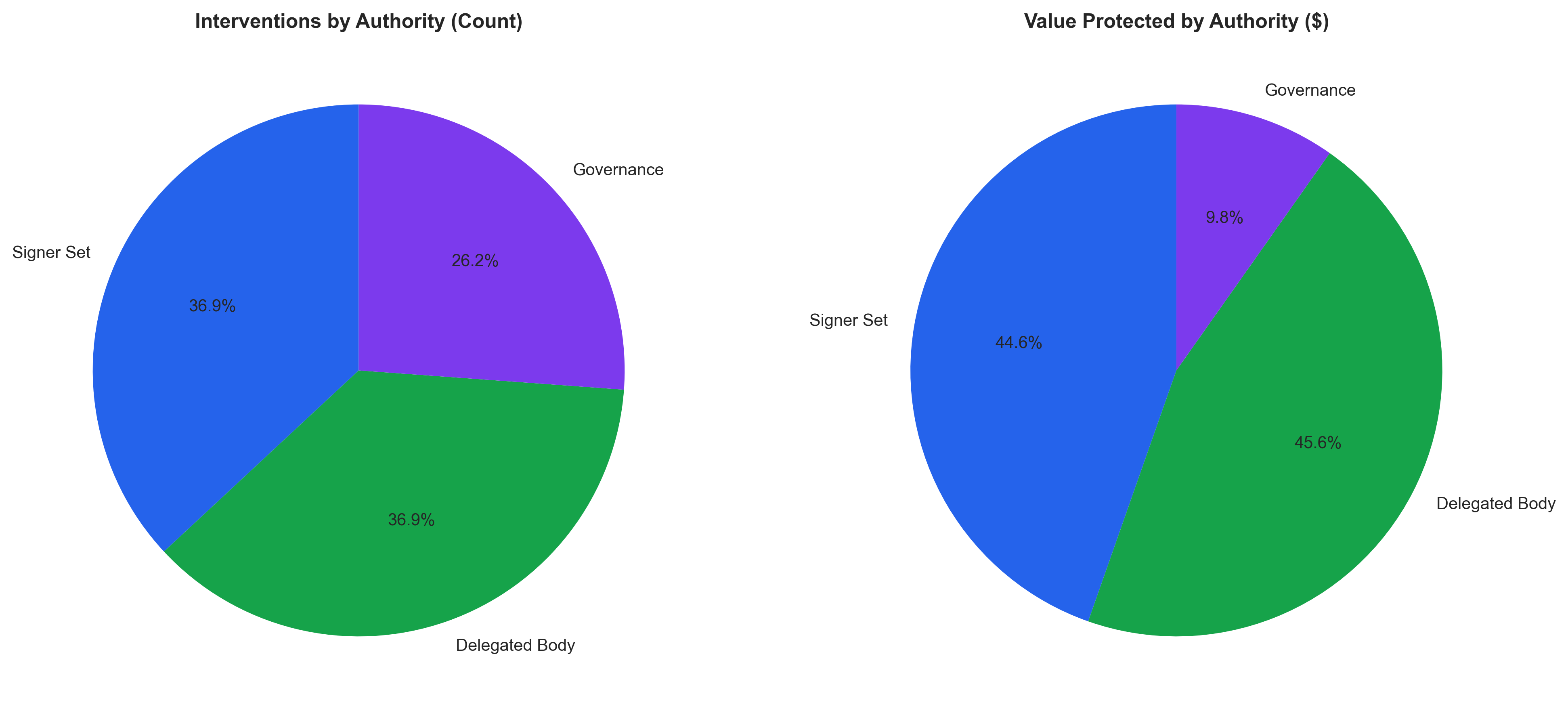}
    \caption{\textbf{Authority Distribution.} Signer Set dominates incident count (executing frequent, smaller interventions), while Governance interventions achieve significant loss prevention through negotiation and recovery of high-value assets. Left: interventions by count. Right: value protected by authority type.}
    \label{fig:authority_dist}
\end{figure}
\fi

\begin{figure}[t]
    \centering
    \includegraphics[width=0.85\linewidth]{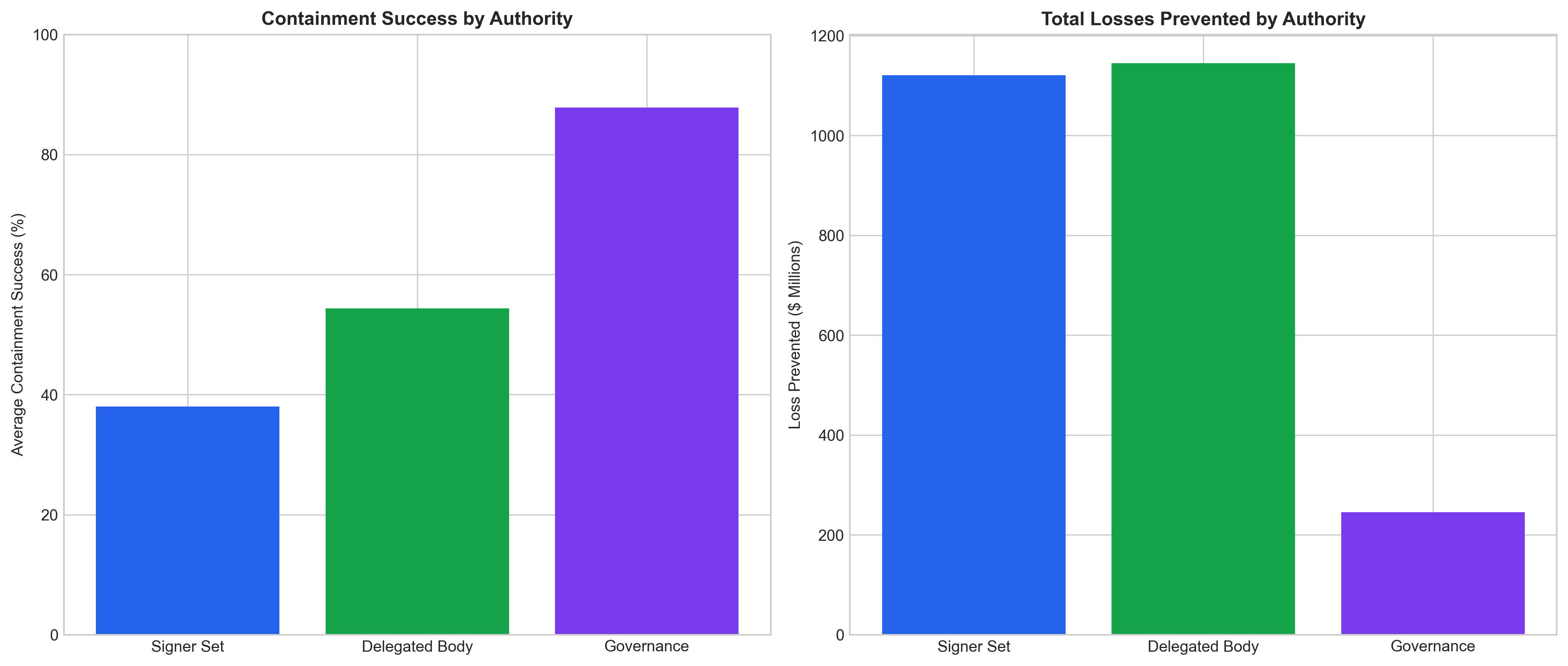}
    \caption{\textbf{Intervention Success Rates.} Comparison of containment success across authority types. In our verified sample, Delegated Body interventions show a higher containment success rate than Signer Set interventions, while Governance-tagged cases represent a small, mixed subset that often includes recovery after earlier containment or offchain negotiation. Right panel shows total losses prevented by authority type.}
    \label{fig:intervention_effectiveness}
\end{figure}

Our intervention incidents data (\Cref{fig:authority_dist}, \Cref{fig:intervention_effectiveness}) confirms the model's Prediction 1: containment time $\Time(m)$ varies systematically by authority type. For the 52 verified intervention incidents:

\begin{itemize}
    \item \textbf{Signer Set} interventions achieve median containment in approximately 30 minutes.
    \item \textbf{Delegated Body} interventions require approximately 60--90 minutes.
    \item \textbf{Governance} interventions, when they occur, operate on timescales of days to weeks (e.g., the Gnosis hard fork required $\approx$30 days).
\end{itemize}

This ordering aligns with intuition: concentrated authority enables faster response, while distributed authority introduces coordination latency.

\ifconcise\else
\subsection{Scope--Authority Matrix: Design Space Population}
\label{subsec:matrix}

\begin{figure}[t]
    \centering
    \includegraphics[width=0.85\linewidth]{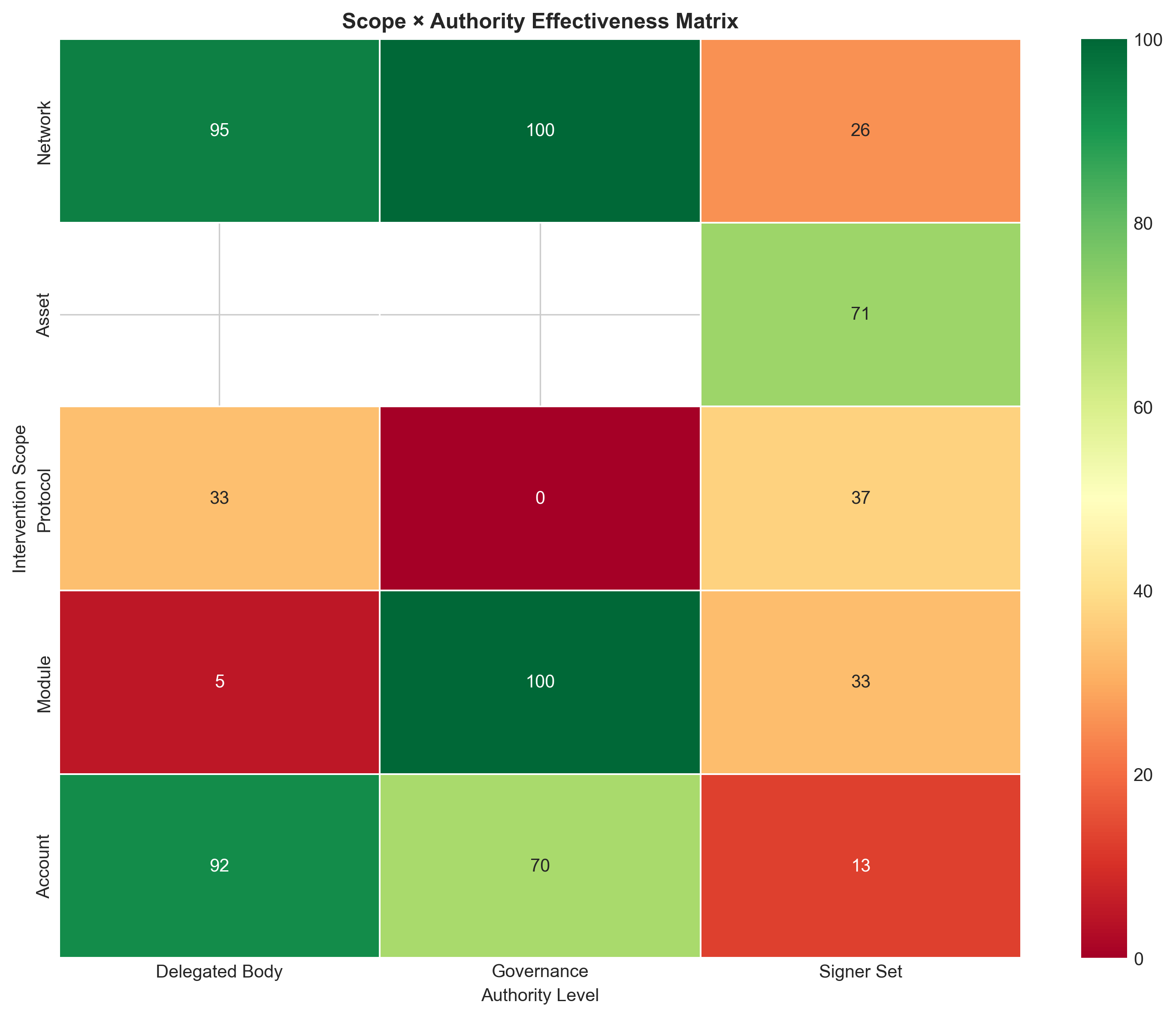}
    \caption{\textbf{Scope $\times$ Authority Heatmap.} Intervention effectiveness (containment success \%) across the taxonomy. Protocol-scope interventions are most frequent, while account-scope actions show high precision.}
    \label{fig:scope_matrix}
\end{figure}

\Cref{fig:scope_matrix} visualizes how real-world interventions populate the Scope $\times$ Authority design space. The data shows that interventions are most frequent at the \emph{Protocol} scope. Notably, the \emph{Protocol/Governance} cell (e.g., MakerDAO Emergency Shutdown) remains largely theoretical or populated by deprecated mechanisms; no major protocol has executed a full governance-triggered shutdown under crisis conditions. While the \emph{Governance} column contains fewer active triggers, it includes high-recovery cases like the Gnosis hard fork and the Ethereum DAO fork. Conversely, the \emph{Protocol/Delegated Body} cell is increasingly populated (e.g., Curve Emergency DAO), reflecting the industry's convergence on delegated safety councils.
\fi

\subsection{Speed--Effectiveness Tradeoff}
\label{subsec:speed-effectiveness}

\begin{figure}[t]
    \centering
    \includegraphics[width=0.9\linewidth]{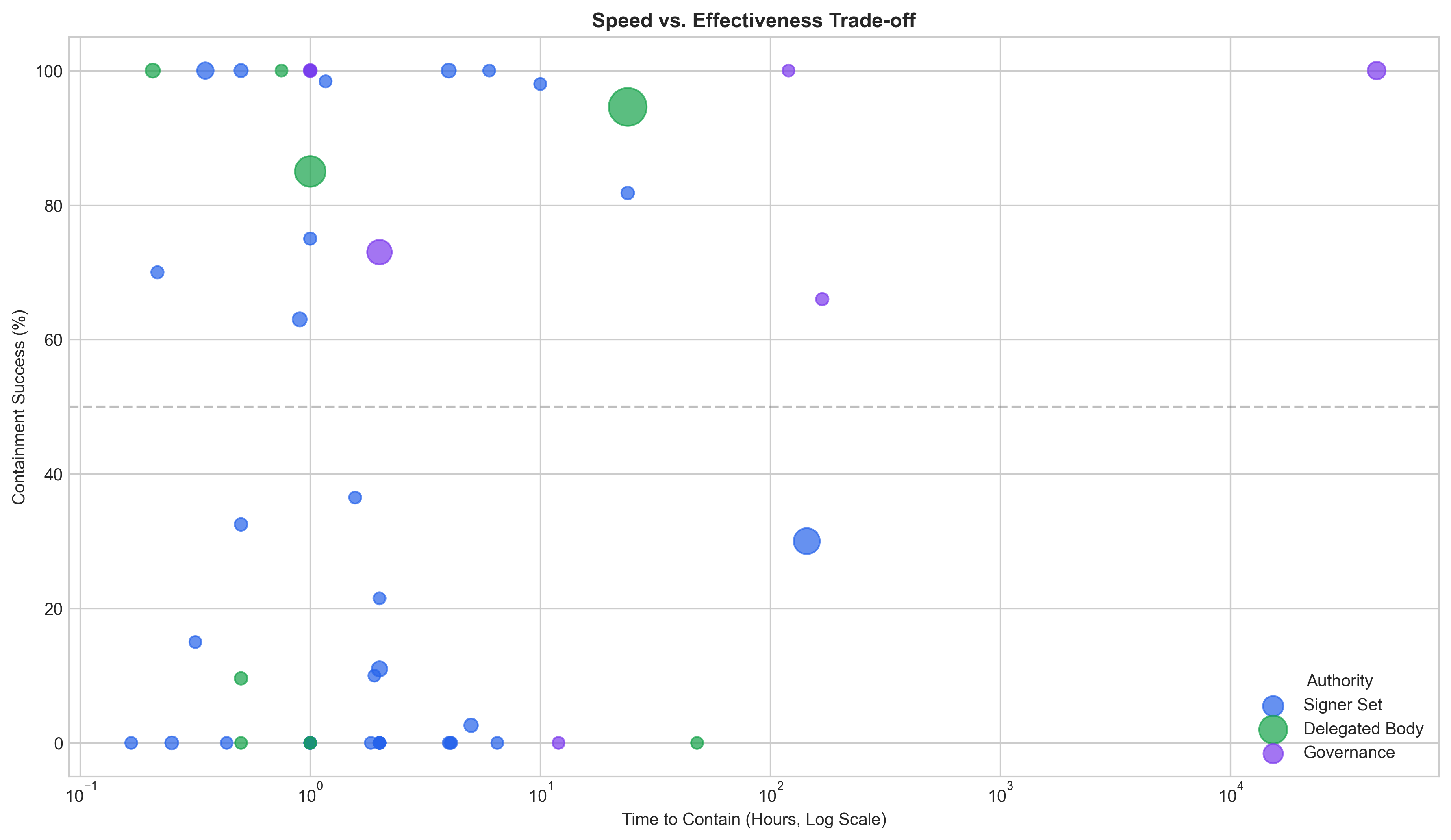}
    \caption{\textbf{Speed-Effectiveness Trade-off.} Relationship between time-to-containment (log scale, hours) and loss prevented. Faster interventions (left side) consistently preserve more value, empirically validating the model's containment term. Bubble size represents loss prevented.}
    \label{fig:calibration_intervention}
\end{figure}

\Cref{fig:calibration_intervention} visualizes the relationship between reaction speed ($\Time(m)$) and containment success for 52 high-fidelity case studies. We define $\Time(m)$ using two observable timestamps: (i) \emph{time-to-detect} = minutes between the first credible alert and the first containment trigger, and (ii) \emph{time-to-contain} = minutes between detection and mechanism execution (pause/freeze/halt). The data supports Prediction 1: faster architectures (like \emph{Signer Set}) significantly minimize loss, though they carry higher centralization costs.

\subsection{Community Sentiment: Calibrating Centralization Cost}
\label{subsec:sentiment}

We collected governance forum discussions (Discourse, Twitter) around intervention incidents and applied an automated sentiment analysis pipeline. For each incident, we extracted $k=20$ representative posts (where available) and processed them using VADER (Valence Aware Dictionary and sEntiment Reasoner) lexicon-based analyzer~\cite{Hutto2014VADER}. VADER is optimized for social media and forum text as it accounts for both lexical features and structural signals such as punctuation and intensifiers. The analyzer returns a normalized \emph{compound} score $s \in [-1, 1]$, where $+1$ is maximally positive and $-1$ is maximally negative.

\ifconcise
The aggregate average sentiment across 271 posts was \textbf{+0.028}, suggesting that in practice, communities tend to accept emergency interventions, though with significant variance.
\else
The aggregate average sentiment across 271 posts was \textbf{+0.028}, suggesting that in practice, communities tend to accept emergency interventions, though with significant variance. We benchmarked seven historical intervention cases plus three general Twitter discussions:

\begin{itemize}
    \item \textbf{Flow (Dec 2025):} 20 posts, avg sentiment +0.167 (supportive -- community acknowledged necessity of intervention)
    \item \textbf{Liqwid (Oct 2025):} 20 posts, avg sentiment +0.204 (supportive -- flash crash response was seen as protective)
    \item \textbf{StakeWise (Nov 2025):} 20 posts, avg sentiment +0.210 (supportive -- multisig recovery welcomed)
    \item \textbf{Anchor (May 2022):} 20 posts, avg sentiment -0.004 (neutral -- mixed reactions to UST crisis response)
    \item \textbf{Gnosis/Balancer (Nov 2025):} 3 posts, avg sentiment -0.034 (neutral -- complex reaction to fork proposal)
    \item \textbf{AAVE (May 2025):} 20 posts, avg sentiment -0.201 (skeptical -- community concerns about intervention scope)
    \item \textbf{BNB (May 2025):} 20 posts, avg sentiment +0.095 (neutral -- mixed community response)
    \item \textbf{DeFi-Hacks-General (Twitter):} 48 posts, avg sentiment -0.267 (skeptical -- general community concerns)
    \item \textbf{Emergency-Pause (Twitter):} 50 posts, avg sentiment -0.128 (skeptical -- concerns about pause mechanisms)
    \item \textbf{Recovery-Actions (Twitter):} 50 posts, avg sentiment +0.236 (supportive -- community favors recovery efforts)
\end{itemize}
\fi

This validates Prediction 3 and our standing centralization cost formulation:
\[
\CentralizationCost(m) = \text{MarketCap} \times \text{DiscountRate}(m) \times (1 - \bar{s})
\]
where $\bar{s} \in [-1, 1]$ is the average sentiment score. Positive sentiment reduces the effective discount rate (community accepts the mechanism), while negative sentiment increases it (community views mechanism as overreach).

\ifconcise
Our 52 verified cases reveal a paradox: Signer Sets dominate volume (38 cases, 73.1\%) but achieve only 38.0\% containment success, while the Governance-coded subset reaches 87.8\% on a much smaller sample (5 cases). That Governance figure should be read cautiously. It is a mixed constitutional/recovery bucket, not a pure instant-containment benchmark. In the revised coding, we exclude Aave v2 and Alpha Homora V2 from Governance because their operative intervention path was guardian-style in the former case and core-team-led in the latter, rather than collective governance. Even among the remaining five cases, some still combine governance authorization with earlier freezes or delayed recovery stages. This reframes the ``Immutability Paradox'' as a \emph{Scope-Authority Matching} problem: protocols fail not because governance is ``slow,'' but because they deploy Direct Democratic mechanisms for Oligarchic tasks (fast containment) while underutilizing them for Constitutional tasks (network recovery).

\else
\subsection{Empirical Support for the Political Analogy: The Speed-Scope-Success Paradox}
\label{subsec:polia-validation}

The empirical performance data strongly supports the political theory analogies proposed in \Cref{sec:taxonomy-authority}.
Our 52 verified intervention cases reveal distinct performance patterns across the authority spectrum:

\begin{itemize}
    \item \textbf{Signer Set (Oligarchy):} Dominates incident volume (38 cases, 73.1\%) and value protected (\$0.55B, 32\% of total), with 38.0\% containment success. This reflects its role as the ``first responder'' for routine protocol, asset, and account-scope interventions where speed trumps deliberation.
    
    \item \textbf{Delegated Body (Representative Democracy):} Handles mid-complexity cases (9 cases, 17.3\%, \$0.88B protected) with 54.4\% success. This intermediate performance reflects the coordination burden of representative structures without the procedural legitimacy of full governance or the operational agility of signer sets.
    
    \item \textbf{Governance (Direct Democracy):} Achieves the highest success rate in our small Governance-coded subset (87.8\%, 5 cases, 9.6\%, \$0.17B protected). Critically, these are predominantly \emph{Network-scope}, recovery-stage, or otherwise hybrid interventions (e.g., Gnosis Hard Fork, Sui Recovery) where deliberative processes enable fundamental protocol state changes that signer sets cannot execute. The ``success'' here is therefore closer to constitutional legitimacy and comprehensive recovery than to immediate exploit containment.
\end{itemize}

This data reframes the ``Immutability Paradox'' as a \emph{Scope-Authority Matching} problem: protocols fail not because governance is ``slow,'' but because they deploy \emph{Direct Democratic} mechanisms for \emph{Oligarchic} tasks (fast containment) while underutilizing them for \emph{Constitutional} tasks (network recovery, history revision), where deliberative processes can be decisive.

\subsection{Detailed Incident Analysis}
\label{subsec:detailed-cases}

For completeness, we provide expanded narratives for representative intervention cases:

\subsubsection{Network-scope interventions}
\label{sec:incidents-network}

\paragraph{Signer Set: BNB Chain and Harmony.}
During the October 2022 BNB Chain bridge exploit (``BSC Token Hub'')~\cite{BNBChainResponse, MerkleScienceBNB2022}, validators coordinated to halt block production as an emergency containment measure. This intervention reduced further drainage but temporarily suspended unrelated activity on the network, making the collateral liveness cost immediately salient. Similar dynamics appeared in Harmony's Horizon Bridge exploit~\cite{Harmony2022Exploit} and later Berachain's validator-coordinated halt~\cite{Berachain2025PostMortem}. The episode illustrates a recurrent legitimacy tension for network-scope actions under concentrated authority: speed and containment are gained at the price of broad disruption and a perception of discretionary control.

\paragraph{Governance: The DAO and Gnosis Forks.}
Following the November 2025 Balancer exploit and subsequent fund freezes, Gnosis Chain executed a governance-approved hard fork (December 2025) to recover a reported \$9.4M in assets that remained frozen onchain~\cite{Gnosis2025HardFork,Gnosis2025BalancerHack}. The episode highlights a characteristic network-scope governance tension: a deliberate, procedurally justified intervention can be perceived as protective, yet it reopens concerns about history revision and precedent.

\subsubsection{Asset-scope interventions}
\label{sec:incidents-asset}

\paragraph{Signer Set: Tether and Circle.}
Asset issuers frequently retain contract-level controls enabling address blocking or freezing. Circle's USDC terms explicitly describe ``Blocked Addresses'' and reserve the ability to freeze USDC associated with such addresses~\cite{CircleUSDCTerms2025}. This authority mode yields high operational responsiveness but embeds discretion at the asset layer, raising legitimacy questions for users who treat stablecoins as credibly neutral settlement assets. In January 2026, Tether executed its largest-ever freeze, blocking \$182M in USDT wallets linked to Venezuelan state oil company PDVSA~\cite{TetherVenezuela2026}.

\paragraph{Delegated Body: Bridge Governance.}
During the November 2025 Balancer exploit, the Gnosis Bridge Governance Board temporarily halted outflows of major tokens (GNO, wstETH, USDC, WETH, and others) to prevent asset drainage~\cite{Gnosis2025BalancerHack}. The Curve Emergency DAO provides another example: while its actions are limited to the protocol-level, its powers are deliberately constrained to asset-specific actions such as stopping CRV emissions on gauges or pausing the Peg Stabilization Reserve while explicitly \emph{not} enabling deposit/withdrawal freezes~\cite{CurveExploitLlamaRisk}. This precision reflects a design choice to limit blast radius: the subDAO can address inflation bugs or peg failures without disrupting the core DEX functionality.

\subsubsection{Protocol-scope interventions}
\label{sec:incidents-protocol}

\paragraph{Signer Set: Admin Controls and Kill Switches.}
The fastest protocol-scope responses rely on concentrated key authority. Liqwid's Proposal 44 (March 2024) explicitly granted the Core Team a single-signature ``kill switch'' to halt market batching within 5--15 minutes for oracle failure scenarios~\cite{Liqwid2024Governance}. The proposal notes that this power coexists with a 4-of-X Pause Guardian multisig (Delegated Body), illustrating defense-in-depth: the kill switch maximizes speed but concentrates trust, while the Guardian adds oversight at the cost of coordination latency.

\paragraph{Delegated Body: Emergency subDAOs.}
Protocols are increasingly converging on bounded delegation as a ``sweet spot.'' Aave distinguishes governance from protocol guardians, who hold time-bounded multisig authority to pause specific markets or the entire protocol~\cite{AaveGuardians,AaveACLManager}. Similarly, Radiant Capital's response to a compromised developer wallet (October 2024) relied on its DAO Council to execute a cross-chain pause on Arbitrum, BSC, and Base. The \$50M exploit via compromised hardware wallets began at 15:46 UTC and was contained by 17:40 UTC across all chains, with compromised signers removed from multisig by 22:10 UTC. While the initial compromise was severe, the Delegated Body structure allowed for a coordinated, legally recognized response that a pure Signer Set might have disorganized and a pure Governance vote would have delayed.

Additionally, Curve Finance's July 2023 \$62M exploit response was managed via its \emph{Emergency DAO}, demonstrating that a delegated committee can act with the speed of a signer set but with greater transparency and role separation~\cite{CurveExploitLlamaRisk}. Latest designs, such as Balancer V3, formalize this further with explicit mandate limits for these delegated bodies~\cite{BalancerV3EmergencySubDAO}.

\paragraph{Governance Process: ESM and Shutdown.}
MakerDAO (now Sky) famously specified an Emergency Shutdown Module (ESM) intended as a last-resort protection mechanism. Shutdown was governed by protocol-defined triggers (via the ESM) and aimed to unwind positions and return collateral to users. However, Sky has since deprecated the ESM mechanism~\cite{MakerDAOEmergencyShutdown}, illustrating that such ``nuclear options'', while theoretically robust, are difficult to maintain due to the game-theoretic risks of malicious triggering and the immense coordination cost of unwinding state.

Similarly, Euler Finance (March 2023) exemplified the ``Protocol/Governance'' tension: lacking an immediate admin override, the protocol was drained of \$197M. Recovery required a high-pressure offchain negotiation (``social layer intervention'') to compel the return of funds~\cite{Euler2023Exploit}, highlighting that pure governance architectures often rely on legal or social backstops when onchain speed is insufficient. We therefore do not treat Euler as a clean example of onchain governance containment. It is better understood as a recovery case in which negotiation dominated the resolution path.

\paragraph{The Balancer Case and Window Expiry.}
The November 2025 Balancer V2 exploit provides a nuanced illustration of protocol-scope safety. Composable Stable Pools (CSP) in V2 were designed with fixed pause windows to ensure eventual immutability. In pools where these windows had expired (CSPv5), the exploit succeeded because emergency pausing was no longer technically possible. Conversely, in CSPv6 pools where pause windows remained active, following \emph{Hypernative} detection of the attack, the team successfully triggered a pause, containing the loss. This contrast highlights a core design trade-off: fixed-length pause windows improve ``liveness'' and perceived immutability but reduce the ability to intervene against vulnerabilities discovered after deployment. Later iterations, such as Balancer's V3 newer architecture, introduced a dedicated emergency governance layer, \emph{Emergency subDAO}, to manage these powers with clearer delegation and mandate limits~\cite{BalancerPostMortem2025,BalancerV3EmergencySubDAO}. The Cork Protocol incident in May 2025, triggered by a Uniswap v4 hook exploit, illustrates how a ``Protocol $\times$ Signer Set'' emergency mode can still mitigate novel threats: after an alert, the team convened an emergency war room with external security partners, SEAL911, and paused remaining markets within roughly an hour, limiting further losses even though the initial exploit had already drained funds from the targeted market~\cite{CorkPostMortem2025}.\footnote{SEAL911 is a security ``hotline'' and rapid-response coordination channel run by members of the Security Alliance (SEAL), where protocols, users, and auditors can raise active or imminent incidents so that experienced responders can help triage, coordinate a war room, and mitigate attacks as they unfold. It operates under the Whitehat Safe Harbor Agreement (SHA), which provides legal protections for good-faith responders assisting in incident mitigation and fund recovery~\cite{seal911github,dedaubseal,piperwhitehat}.}

\subsubsection{Module-scope interventions}
\label{sec:incidents-module}

\paragraph{Delegated Body: Feature-Specific Pauses.}
Module-scope interventions restrict a specific function (e.g., liquidations) while allowing other protocol activity to continue. Aave's role-based control system explicitly supports pausing at the pool or reserve level by emergency administrators, illustrating the general pattern of narrowing blast radius via feature-specific switches~\cite{AaveACLManager,AaveDPIFreezeProposal}.

\paragraph{Governance Process: Function Toggles.}
Some protocols route feature disabling or emergency patches through governance, trading response time for broader legitimacy.
In our framework, these episodes populate the module-scope / governance cell when the intervention is limited to a component but procedurally collective.
MakerDAO's emergency governance proposal to adjust risk and governance parameters for USDC‑PSM during the March 2023 USDC depeg exemplifies this pattern~\cite{MakerDAOEmergency2023}, as does dYdX's use of margin and position‑size adjustments in the YFI‑USD market as a de‑facto circuit‑breaker during the November 2023 liquidation‑stress incident~\cite{dYdXSushiYFI}.

\subsubsection{Account-scope interventions}
\label{sec:incidents-account}

Account-scope interventions represent the highest precision in our taxonomy: they target specific addresses or balances without disrupting the broader system.
This surgical approach minimizes collateral disruption but requires robust evidence and instrumentation.

\paragraph{Signer Set: Key-Based Targeted Restrictions.}
Asset issuers and protocol administrators frequently retain the ability to freeze or quarantine specific addresses. Tether and Circle maintain address blacklists that can be updated unilaterally by the issuer, enabling rapid response to sanctions compliance or exploit containment. During the November 2025 Balancer exploit, Sonic Labs deployed its \texttt{freezeAccount} mechanism within two hours to freeze suspected attacker addresses~\cite{Sonic2025Freeze}. However, the Sonic/Beets incident uncovered a critical limitation: while the freeze blocked direct transfers, 78.5\% of the funds were lost because the attacker used \texttt{permit()} signatures to approve transfers from a different, unfrozen address. This signature-based bypass illustrates that Account-scope interventions must block not only transactions but also state-changing signatures to be fully effective. Similarly, after the March 2022 Ronin Bridge exploit (\$625M), attacker addresses were identified and partially frozen, though the delayed detection (6 days) limited recovery~\cite{Ronin2022Exploit}. These cases illustrate the speed advantage of Signer Set authority at the Account scope, though they embed significant discretionary power.

\paragraph{Delegated Body: Council-Authorized Remediation.}
In the December 2025 Flow incident, validators halted the network after counterfeit tokens were created and approximately \$3.9M was extracted. Flow's post-mortem emphasizes containment without rollback, describing an ``Isolated Recovery'' approach that preserved legitimate user activity while restricting only 1,060 addresses (under 0.01\% of total accounts) implicated in the exploit. The Community Governance Council (CGC), operating under validator-authorized boundaries, executed token burns to neutralize counterfeit assets~\cite{FlowPostmortem2026,Flow2025Recovery}. This episode demonstrates that network-layer exploits can be remediated at the account level when proper instrumentation exists.

The November 2025 StakeWise recovery provides a similar case at the application layer. Following the Balancer V2 exploit, the StakeWise emergency multisig (a 7-stakeholder body with the core team as only one signer) used the protocol's ``controller'' role to burn osETH and osGNO tokens in the exploiter's wallet and re-mint them to a DAO-controlled address~\cite{StakeWise2025Recovery}. By manipulating state at the account level, the multisig recovered \$20.7M without disrupting stable users. The episode also illustrates the lifecycle of emergency powers: after successful intervention, the team initiated a governance vote to renounce these capabilities, trading future response capability for reduced standing centralization cost.

Similar coordination occurred in the November~30, 2025 Yearn yETH exploit. To recover assets, Yearn worked with the pxETH issuer (Redacted Cartel) to burn 857.49 pxETH ($\approx$ \$2.4M) from the attacker's wallet and re-mint them to protocol control~\cite{Yearn2025Recovery}. This incident demonstrates how asset-issuer authority can serve as a surgical remediation layer when protocol-level pauses are absent or insufficient.

\paragraph{Governance Process: Stake-Weighted Vote.}
The May 2025 Cetus exploit on Sui (\$220M stolen) produced the most rigorous governance-authorized account intervention to date. After validators initially froze \$162M by refusing to process transactions from two attacker addresses (a Delegated Body action), Cetus called for a community vote to authorize a protocol upgrade that would recover the frozen funds. The Sui Foundation abstained to maintain neutrality, and the vote concluded with \textbf{90.9\% of stake} voting ``Yes''~\cite{SuiCetusVote,UnchainedCetus2025}. The recovered funds were transferred to a 4-of-6 multisig (Cetus, Sui Foundation, OtterSec) for distribution. This is a hybrid case. The delegated freeze provided the first containment, while governance supplied the explicit mandate for recovery and redistribution.

\paragraph{The VeChain Classification Dispute.}
The distinction between ``admin freeze'' and ``governance-led blocklist'' is not merely semantic but central to legitimacy. Following Bybit's categorization of VeChain as having ``hidden freezing capabilities''~\cite{ByBit2025SecurityReport}, VeChain publicly refuted this, clarifying that their mechanism is a validator-enforced blocklist authorized by community governance (originally voted in December 2019 following a theft), rather than a unilateral admin key~\cite{VeChain2025Refutation}. This dispute highlights the \emph{Authority} axis of our taxonomy: two mechanisms may achieve the same \emph{Scope} (account freezing) but differ fundamentally in their \emph{Authority} source (Signer Set vs. Governance), drastically altering their perceived centralization cost.

\subsection{Summary: Theory-Empirics Alignment}
\label{subsec:validation-summary}

Our empirical analysis supports all three predictions of the expected cost model, though the second relies on a proxy operationalization rather than a direct market-wide measure of collateral disruption:
\begin{enumerate}
    \item \textbf{Containment time varies by authority:} Signer Set (30 min) $<$ Delegated Body (60--90 min) $<$ Governance (days), validating $\Time(m)$.
    \item \textbf{Losses follow power law:} 80/20 concentration validates focusing intervention capability on rare, catastrophic events.
    \item \textbf{Sentiment modulates centralization cost:} Positive community acceptance reduces effective standing costs, validating the culture multiplier.
    \item \textbf{Narrower scope does not appear to sacrifice containment:} Using scope breadth as a proxy for blast potential, Account/Module interventions outperform Protocol/Network interventions on median containment success (25.8\% vs.\ 10.0\%), show slightly higher mean containment success (43.5\% vs.\ 40.4\%), and achieve a slightly faster median containment time (90 vs.\ 114 minutes). This is a proxy test, not a direct estimate of market-wide collateral loss, but it is consistent with Prediction 2 rather than contradicting it.
\end{enumerate}

\fi

\section{Design Implications}
\label{sec:implications}

Our theoretical model (\Cref{sec:theory}) and empirical findings (\Cref{sec:empirical}) jointly suggest several design principles for emergency governance in decentralized protocols.

\subsection{The Delegation Sweet Spot}
\label{subsec:delegation}

The expected cost function $\ExpectedCost(m)$ reveals a non-monotonic relationship between authority concentration and total social cost. Our empirical Speed-Scope-Success Paradox data validates this theoretical prediction:

\begin{itemize}
    \item \textbf{Pure governance} ($m_{\text{gov}}$) minimizes standing centralization cost but maximizes containment time, making it unsuitable for fast-moving exploits. In our current sample, the Governance-coded subset reaches 87.8\% success, but that figure comes from a very small, mixed set of recovery-stage and hybrid cases rather than a pure containment benchmark.
    \item \textbf{Signer set} ($m_{\text{key}}$) minimizes containment time but imposes a large standing trust tax, reducing protocol valuation even absent any incident. Empirically, this dominates volume (73.1\% of cases) but achieves only 38.0\% containment success.
    \item \textbf{Delegated body} ($m_{\text{council}}$) occupies the empirical sweet spot: 54.4\% success rate on 17.3\% of cases with \$0.88B protected, reflecting bounded authority with faster-than-governance response but without the Signer Set's trust tax.
\end{itemize}

The emergence of \emph{Emergency subDAOs} (pioneered by Curve, adopted by Balancer V3, and variants like Aave's Protocol Guardians) reflects practitioner convergence toward this sweet spot.
These bodies operate under explicit mandates, time-bounded powers, and reporting requirements, features that reduce $\CentralizationCost(m)$ while preserving containment speed.

\begin{remark}[Legitimacy extends beyond a cost term]
Several cases in our dataset suggest that legitimacy emerges actively through process and post-crisis conduct, extending beyond ex ante mechanism choice. StakeWise's later renunciation vote and the Sui Foundation's abstention both matter because they shape how intervention authority is interpreted after the fact. The cost model captures part of this through sentiment, though it serves as a partial account of legitimacy formation rather than a comprehensive one.
\end{remark}

\subsection{Scope Matters: Precision Reduces Blast Radius}
\label{subsec:scope-implications}

Our taxonomy distinguishes five levels of scope, from network-wide halts to account-level freezes.

The empirical data supports a narrower version of the blast-radius claim than we originally stated. Using scope breadth as a proxy for blast potential, Account/Module interventions do not underperform Protocol/Network interventions on containment outcomes. In the 52-case high-fidelity subset, narrower interventions show slightly higher mean containment success (43.5\% vs.\ 40.4\%), materially higher median containment success (25.8\% vs.\ 10.0\%), and slightly faster median containment time (90 vs.\ 114 minutes). This is not a direct measure of total collateral disruption, but it is consistent with the design intuition that more precise interventions can preserve response quality without requiring broader shutdowns.

For protocol designers, this implies:

\begin{itemize}
    \item Invest in \emph{instrumentation} that enables targeted response (e.g., per-account freeze hooks, module-level circuit breakers).
    \item Avoid reliance on ``nuclear options'' (full chain halts) except as a last resort.
    \item Design upgrade paths that \emph{increase} precision over time as tooling matures.
\end{itemize}

\subsection{The Culture Multiplier}
\label{subsec:culture}

Blast radius cost is not uniform across ecosystems.
Chains with strong ``DeFi-maxi'' or permissionless cultures suffer disproportionate reputational damage from interventions, even successful ones.

Conversely, chains targeting regulated use cases (RWA, payments, institutional custody) may experience a \emph{regulatory premium}: the presence of robust override capability increases rather than decreases valuation.

Designers should calibrate their mechanism choice to community expectations.

A ``culture multiplier'' $\gamma$ can be incorporated into the model:
\[
\BlastRate(m) = \gamma \cdot \text{Scope\%} \cdot \frac{\text{Daily Volume}}{1440}
\]

where $\gamma$ is high for permissionless chains and low for compliance-oriented chains.

\ifconcise\else
\subsection{Sunset Clauses and Ossification}
\label{subsec:sunset}

The Balancer CSPv5 case illustrates the risk of \emph{premature ossification}: pause windows that expire before all code paths are proven safe.

We recommend:
\begin{itemize}
    \item \textbf{Conditional sunsets}: Pause capability should expire only after explicit security milestones (e.g., formal verification, extended bug-bounty periods without critical findings).
    \item \textbf{Renewable mandates}: Delegated bodies should require periodic re-authorization, preventing indefinite entrenchment.
\end{itemize}
\fi

\subsection{A Decision Framework}
\label{subsec:framework}

To make the theoretical model actionable, we map observable protocol and exploit parameters to the formal model variables in \Cref{tab:risk_factors}. This allows designers to calibrate the calculator tool using evidence-based heuristics.

\begin{table}[h]
\centering
\caption{\textbf{Mapping Protocol Parameters to Model Variables.} This table standardizes the mapping between observable protocol and exploit characteristics and formal variables in our expected cost model.}
\label{tab:risk_factors}
\small
\begin{tabularx}{\linewidth}{@{}>{\raggedright\arraybackslash}p{2.3cm}>{\raggedright\arraybackslash}p{2.8cm}>{\raggedright\arraybackslash}p{3.8cm}X@{}}
\toprule
\textbf{Parameter} & \textbf{Relevance} & \textbf{Model Variable} & \textbf{Short Explanation} \\
\midrule
Protocol Type & Asset risk profile & $\Prob[h]$, $\DamageRate(h)$ & AMM vs Lending vs Bridge risk profiles \\
Exploit Type & Containment urgency & $\DamageRate(h)$ & Flash loan (fast) vs Reentrancy (medium) \\
Exploit Novelty & Variant vs. Zero-day & $\Prob[h]$, $\DamageRate(h)$ & Zero-day = max damage rate \\
Audit Status & Preventive security health & $\Prob[h]$ & Weak proxy only; may reduce some known-class risks but does not capture zero-days or socio-technical failures \\
Community Sentiment & Political legitimacy & $\CentralizationCost(m)$, $\BlastRate(m)$ & Trust reduces political cost \\
TVL Affected & Economic scale at risk & $\BlastRate(m)$ & Higher TVL = larger blast radius \\
Security Claims & Breach accountability & $\CentralizationCost(m)$ & ``Immutable'' = high trust tax \\
\bottomrule
\end{tabularx}
\end{table}

\ifconcise\else
\begin{remark}[Practitioner Alignment]
The parameters in \Cref{tab:risk_factors} align closely with factors independently identified by practitioners.
For example, in the Gnosis Community AMA (January~7, 2026), Gnosis co-founder Friederike Ernst outlined analogous decision criteria such exploit type, protocol type, exploit novelty, security claims, audit status, and percentage of chain TVL affected, as factors the core team may evaluate if considering potential intervention.
In a subsequent contribution to the GnosisDAO governance forum~\cite{GnosisForumFramework2026}, we formalized these considerations into a structured decision framework, recommending (i)~pre-defined thresholds for each parameter documented in the governance framework, (ii)~weighted scoring to combine multiple factors into a transparent decision matrix, and (iii)~publication of these criteria before any incident occurs, so that both the community and potential interveners share a common reference point.
This convergence between our theoretical model, practitioner intuition, and the proposed governance framework supports the actionability of the parameter space identified here.
\end{remark}
\fi

\begin{remark}[Scope limits of the cost model]
Two limits matter for interpretation. First, variables such as $\CentralizationCost(m)$ and $\BlastRate(m)$ are best treated as structured thinking tools, not as directly observable quantities that a protocol designer can estimate with high confidence. Second, some influential resolution channels remain partly outside the model. Negotiation, legal pressure, attacker cooperation, and law-enforcement action may dominate the outcome in cases such as Euler or Ronin. This paper treats those channels as external complements or competitors to onchain intervention rather than as fully modeled mechanism choices.
\end{remark}

We propose that protocol designers use our stochastic model as a decision support tool:

\begin{enumerate}
    \item \textbf{Estimate threat parameters}: Probability distribution $\Prob[h]$ and damage rates $\DamageRate(h)$ for plausible exploit scenarios.
    \item \textbf{Estimate mechanism costs}: Standing centralization cost $\CentralizationCost(m)$, containment time $\Time(m)$, and blast rate $\BlastRate(m)$ for candidate architectures.
    \item \textbf{Minimize expected cost}: Choose the architecture $m^*$ that minimizes $\ExpectedCost(m)$ given the protocol's risk profile and community culture.
\end{enumerate}

This framework moves emergency governance from ideology (``decentralization good, admin keys bad'') to quantitative cost-benefit analysis.

\begin{remark}[Audit-status limitation]
Audit status is only a coarse proxy for exploit probability. Audits are scoped to known vulnerability classes, and auditor incentives are not aligned with honest probability estimation. Multiple incidents in our own dataset, including compromised developer hardware and signature-bypass cases, fall outside standard audit coverage by construction. Any use of audit status in the calculator should therefore be treated as a heuristic feature, not as a calibrated probability estimate.
\end{remark}

\section{Conclusion}
\label{sec:conclusion}

Emergency override mechanisms are a pervasive yet under-theorized feature of decentralized protocols.
This paper has offered three complementary contributions toward filling that gap.
First, we introduced a compact \emph{Scope $\times$ Authority} taxonomy that organizes the heterogeneous landscape of emergency mechanisms into a single, two-dimensional design space, and we populated it with prominent real-world episodes spanning chain-level halts, asset freezes, protocol pauses, module toggles, and account-level quarantines.
Second, we formalized the underlying design tradeoff as a stochastic decision-support framework that balances standing centralization cost, containment speed, and collateral disruption, and from it derived three empirical hypotheses.
Third, we assessed this framework against 705 documented exploit incidents (2016--2026), finding that containment time varies systematically by authority type, that losses follow a power-law distribution concentrating risk in rare ``super-hacks,'' and that community sentiment measurably modulates the effective cost of intervention capability. We also find modest empirical support for the scope-blast hypothesis when scope breadth is used as a proxy for blast potential: narrower interventions do not appear to sacrifice containment performance and are slightly faster at the median.
Together, these results reframe emergency governance as a quantitative engineering problem rather than an ideological binary, and provide actionable guidance on delegation sweet spots, precision instrumentation, culture-aware calibration, and conditional sunset clauses for protocol designers navigating the immutability--intervention paradox.

\ifconcise\else
\section{Future Directions}
\label{subsec:future}

We outline directions for future research:

\begin{itemize}
\item \textbf{Formalizing Heterogeneous Stakeholder Costs.}
Different stakeholders (token holders, LPs, integrators, validators) bear asymmetric costs from both exploits and interventions. Formalizing these heterogeneous preferences could yield more nuanced design recommendations.

\item \textbf{Decision Support Tooling.}
The expected cost framework naturally suggests a ``calculator'' tool for protocol designers. Given estimates of threat probabilities and mechanism costs, such a tool could rank candidate architectures and perform sensitivity analysis. We have implemented an open-source prototype of this calculator as part of this research, allowing real-time calculation and sentiment-based calibration of mechanisms.

\item \textbf{Extended Taxonomy Development.}
Our 5$\times$3 framework captures reactive interventions effectively, but \Cref{subsec:pre-execution-prevention-primitives} (e.g., Phylax Credible Layer) suggest need for additional dimensions. Future work could develop a comprehensive taxonomy that includes timing dimensions (pre-execution vs. reactive) and enforcement mechanisms (assertion-based vs. pause-based).

\item \textbf{Cross-Chain Coordination Protocols.}
As protocols operate across multiple chains, emergency response mechanisms face new coordination challenges. Research into standardized cross-chain emergency protocols could enable synchronized responses while preserving chain sovereignty.

\item \textbf{Dynamic Mechanism Selection.}
The expected cost framework assumes static mechanism selection, but real-world conditions may require adaptive approaches. Investigating dynamic mechanism selection based on real-time threat assessment could improve responsiveness.
\end{itemize}
\fi

\ifconcise
\clearpage
\appendix

\section{Supplementary Figures}
\label{app:figures}

\begin{figure}[h]
    \centering
    \includegraphics[width=0.9\linewidth]{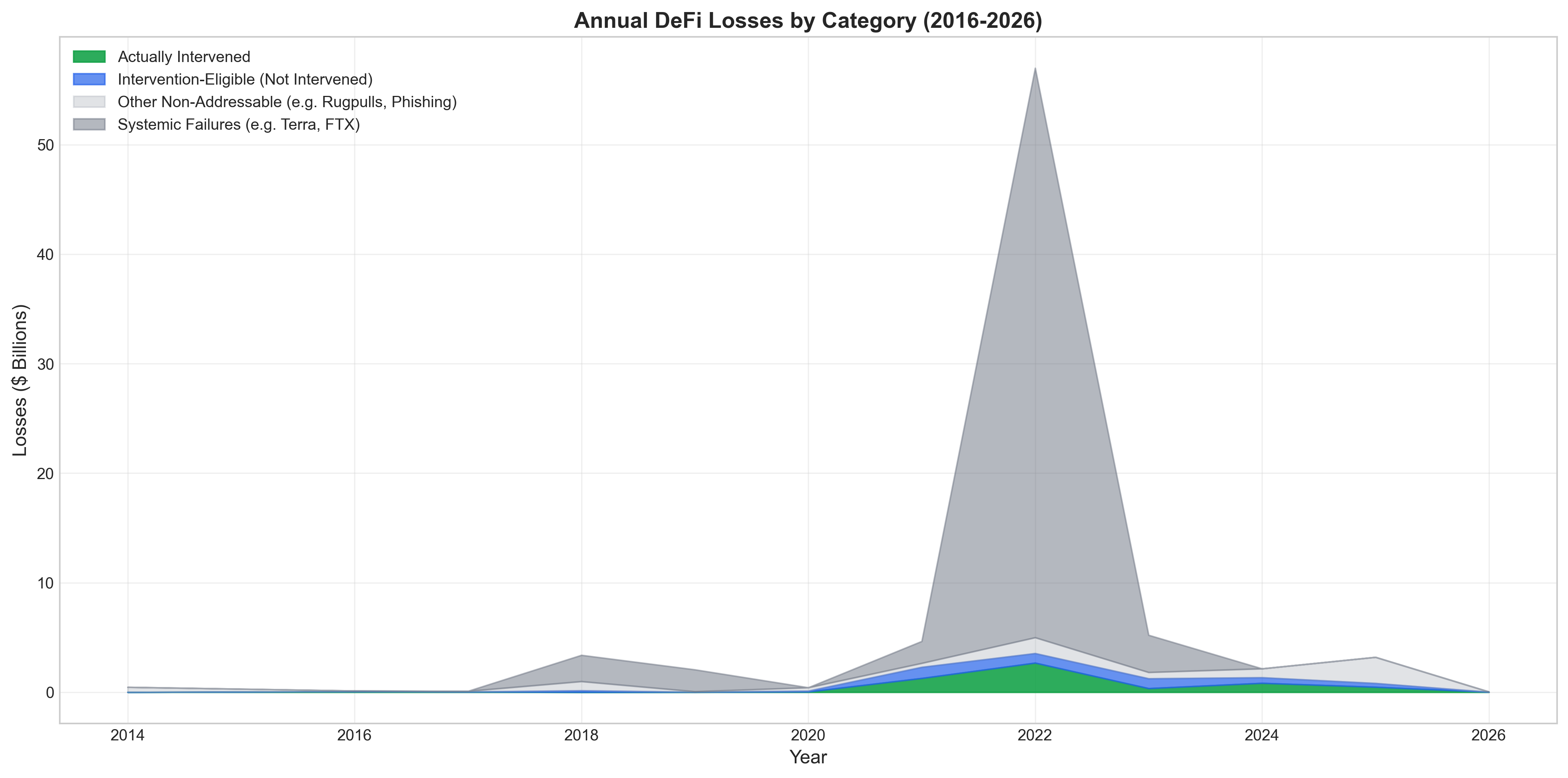}
    \caption{\textbf{Stratification of Losses (2016-2026).} We stratify losses into four layers: \textbf{Systemic Failures} (dark grey, e.g., Terra), \textbf{Other Non-Addressable} (light grey, e.g., rug pulls), \textbf{Intervention-Eligible} (blue), and \textbf{Actually Intervened} (green). This reveals that while systemic events dominate 2022, addressable technical exploits represent a consistent baseline of risk.}
    \label{fig:timeline_stratification}
\end{figure}

\begin{figure}[h]
    \centering
    \includegraphics[width=0.85\linewidth]{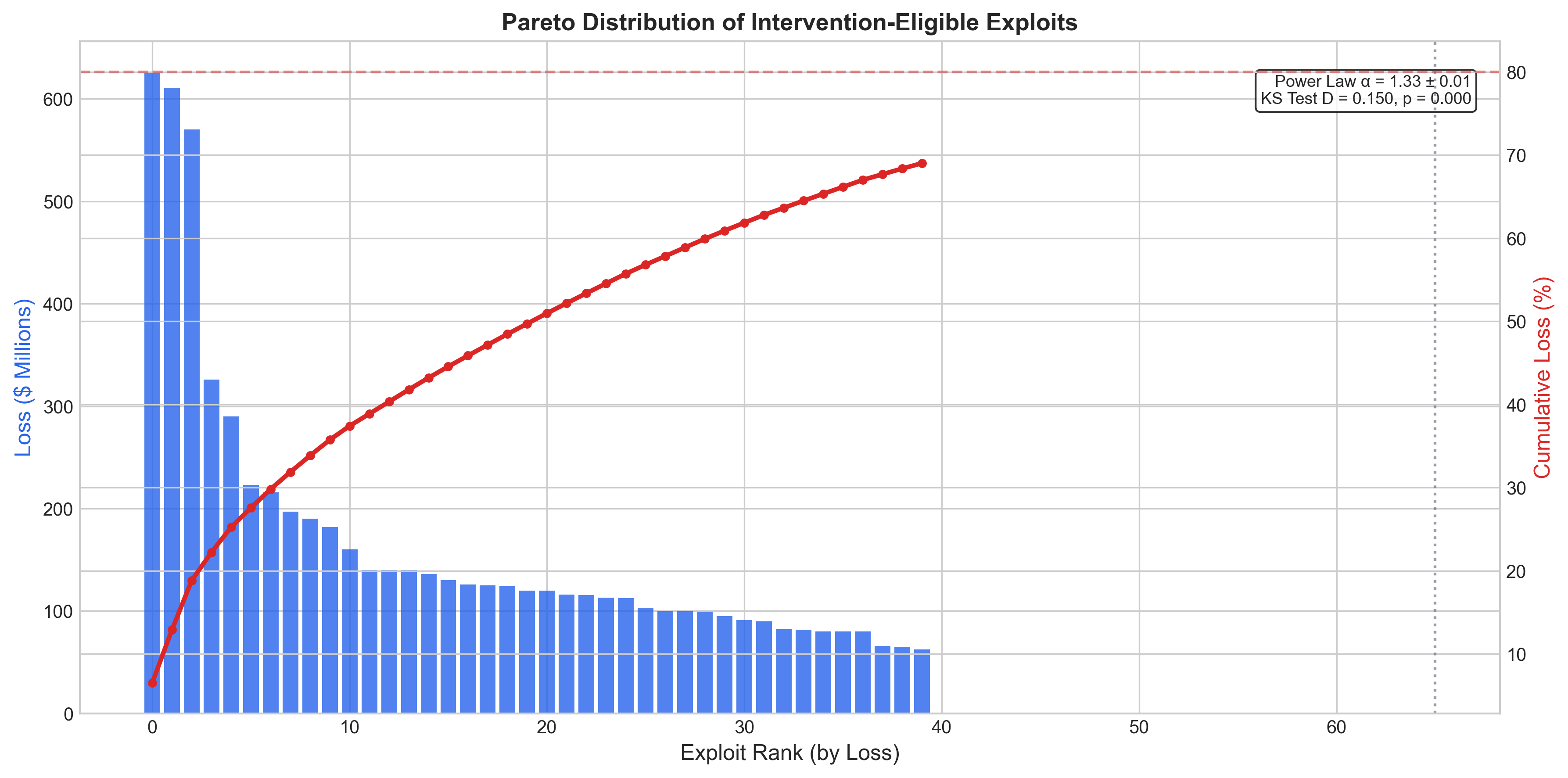}
    \caption{\textbf{Pareto Distribution of Intervention-Eligible Losses.} Approximately 80\% of cumulative losses in our addressable dataset are attributable to fewer than 50 incidents. Power law fit: $\alpha \approx 1.33$, KS test $D=0.150$, $p < 0.001$.}
    \label{fig:pareto_loss}
\end{figure}

\begin{figure}[h]
    \centering
    \includegraphics[width=0.85\linewidth]{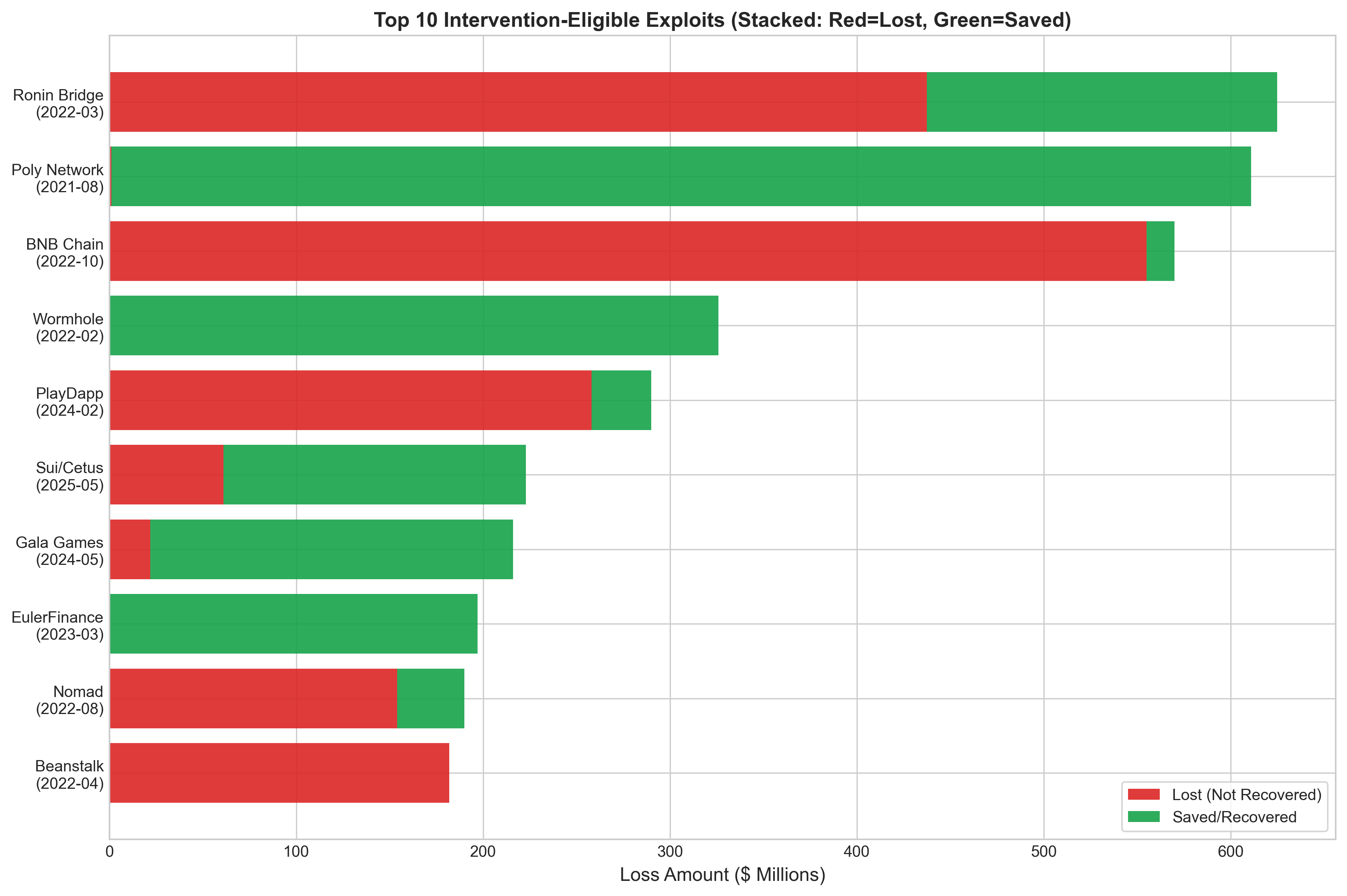}
    \caption{\textbf{Top 10 Intervention-Eligible Exploits.} Stacked bars show losses prevented (green) versus lost (red).}
    \label{fig:top_10_exploits}
\end{figure}

\begin{figure}[h]
    \centering
    \includegraphics[width=0.85\linewidth]{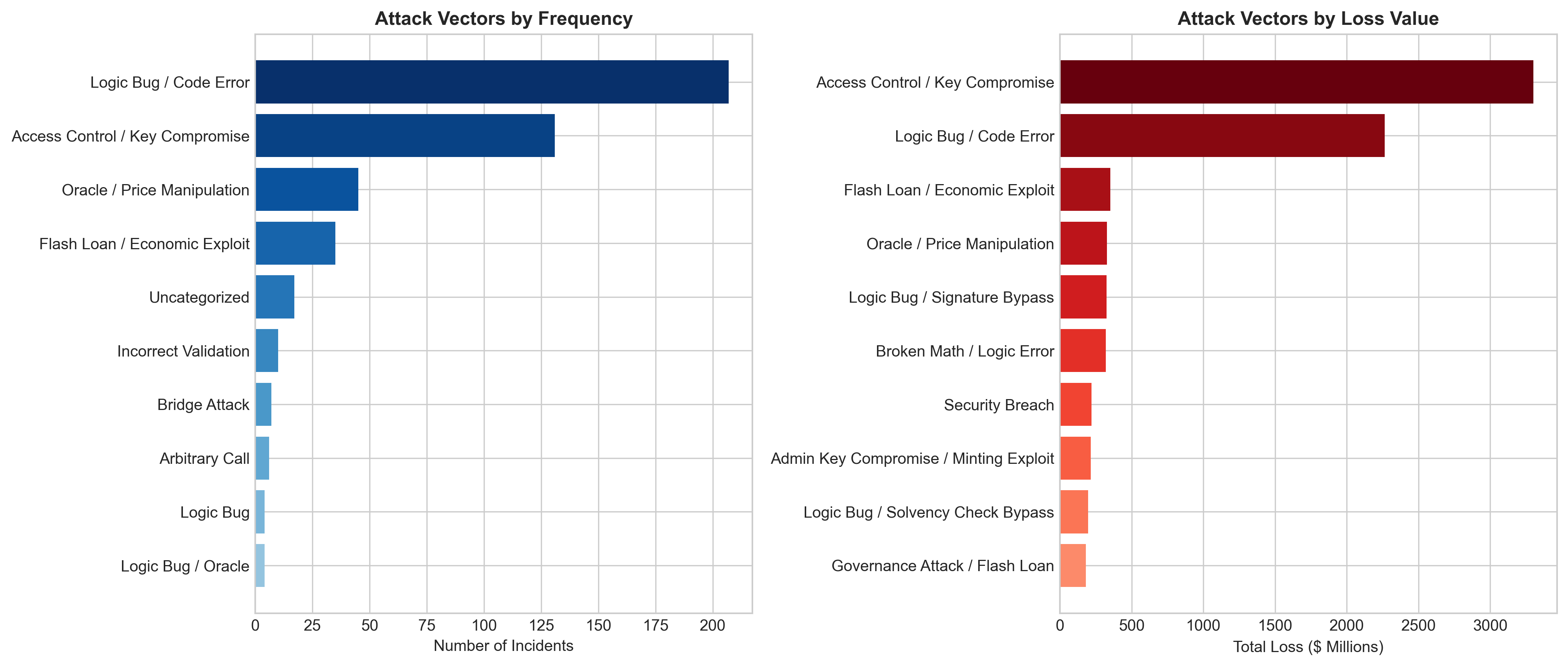}
    \caption{\textbf{Attack Vector Distribution.} While `Logic Errors' and `Access Control' issues are frequent and account for significant losses; complex `Oracle Manipulation' and `Flash Loan' attacks often result in the highest severity incidents.}
    \label{fig:attack_vectors}
\end{figure}

\begin{figure}[h]
    \centering
    \includegraphics[width=0.85\linewidth]{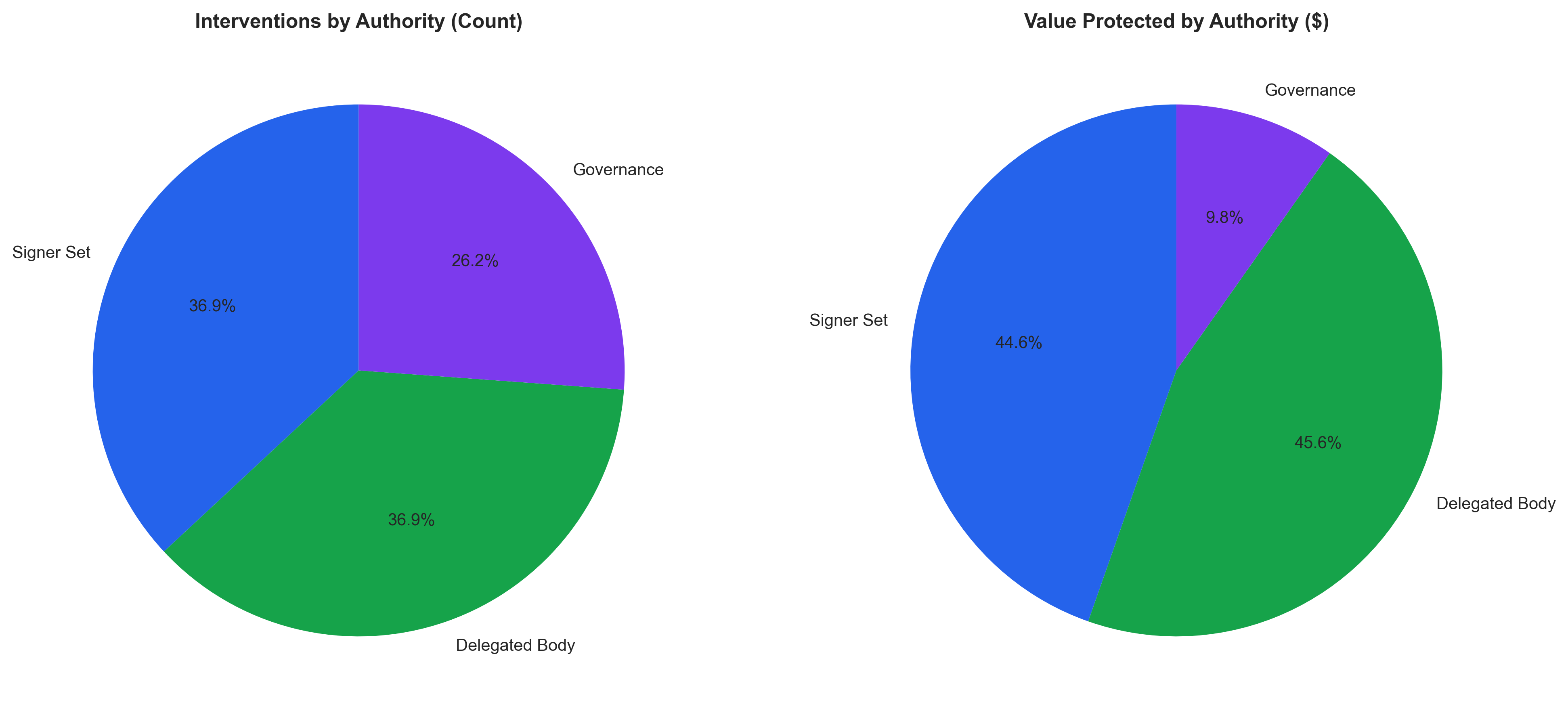}
    \caption{\textbf{Authority Distribution.} Signer Set dominates incident count, while Governance interventions achieve significant loss prevention through negotiation and recovery.}
    \label{fig:authority_dist}
\end{figure}

\begin{figure}[h]
    \centering
    \includegraphics[width=0.85\linewidth]{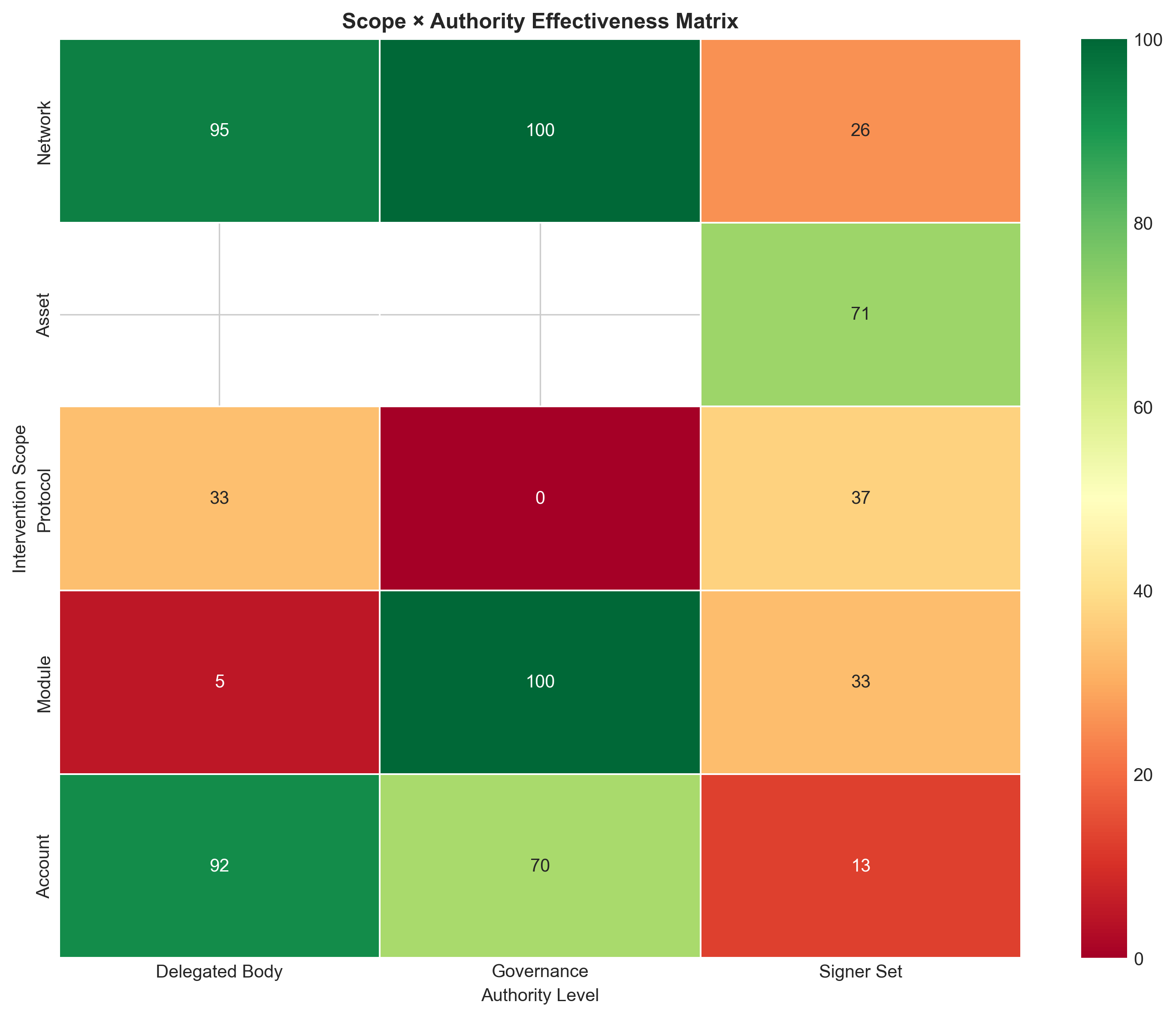}
    \caption{\textbf{Scope $\times$ Authority Heatmap.} Intervention effectiveness (containment success \%) across the taxonomy. Protocol-scope interventions are most frequent, while account-scope actions show high precision.}
    \label{fig:scope_matrix}
\end{figure}

\section{Detailed Incident Analysis}
\label{subsec:detailed-cases}

We provide expanded narratives for representative intervention cases across the Scope $\times$ Authority taxonomy.

\subsection{Network-scope interventions}
\label{sec:incidents-network}

\paragraph{Signer Set: BNB Chain and Harmony.}
During the October 2022 BNB Chain bridge exploit (``BSC Token Hub'')~\cite{BNBChainResponse, MerkleScienceBNB2022}, validators coordinated to halt block production as an emergency containment measure. This intervention reduced further drainage but temporarily suspended unrelated activity on the network, making the collateral liveness cost immediately salient. Similar dynamics appeared in Harmony's Horizon Bridge exploit~\cite{Harmony2022Exploit} and later Berachain's validator-coordinated halt~\cite{Berachain2025PostMortem}.

\paragraph{Governance: The DAO and Gnosis Forks.}
Following the November 2025 Balancer exploit and subsequent fund freezes, Gnosis Chain executed a governance-approved hard fork (December 2025) to recover a reported \$9.4M in assets that remained frozen onchain~\cite{Gnosis2025HardFork,Gnosis2025BalancerHack}.

\subsection{Asset-scope interventions}
\label{sec:incidents-asset}

\paragraph{Signer Set: Tether and Circle.}
Circle's USDC terms explicitly describe ``Blocked Addresses'' and reserve the ability to freeze USDC~\cite{CircleUSDCTerms2025}. In January 2026, Tether executed its largest-ever freeze, blocking \$182M in USDT wallets~\cite{TetherVenezuela2026}.

\paragraph{Delegated Body: Bridge Governance.}
During the November 2025 Balancer exploit, the Gnosis Bridge Governance Board temporarily halted outflows of major tokens~\cite{Gnosis2025BalancerHack}. The Curve Emergency DAO provides another example with powers deliberately constrained to asset-specific actions~\cite{CurveExploitLlamaRisk}.

\subsection{Protocol-scope interventions}
\label{sec:incidents-protocol}

\paragraph{Signer Set: Admin Controls and Kill Switches.}
Liqwid's Proposal 44 (March 2024) explicitly granted the Core Team a single-signature ``kill switch'' to halt market batching within 5--15 minutes~\cite{Liqwid2024Governance}.

\paragraph{Delegated Body: Emergency subDAOs.}
Aave distinguishes governance from protocol guardians who hold time-bounded multisig authority~\cite{AaveGuardians,AaveACLManager}. Curve Finance's July 2023 \$62M exploit response was managed via its Emergency DAO~\cite{CurveExploitLlamaRisk}. Balancer V3 formalizes this further with explicit mandate limits~\cite{BalancerV3EmergencySubDAO}.

\paragraph{Governance Process: ESM and Shutdown.}
MakerDAO's ESM mechanism has since been deprecated~\cite{MakerDAOEmergencyShutdown}. Euler Finance (March 2023) exemplified the ``Protocol/Governance'' tension: lacking an immediate admin override, recovery required offchain negotiation~\cite{Euler2023Exploit}.

\paragraph{The Balancer Case and Window Expiry.}
The November 2025 Balancer V2 exploit highlights a core design trade-off: fixed-length pause windows improve immutability but reduce intervention capability~\cite{BalancerPostMortem2025,BalancerV3EmergencySubDAO}. The Cork Protocol incident (May 2025) illustrates emergency war room coordination with SEAL911~\cite{CorkPostMortem2025}.\footnote{SEAL911 operates under the Whitehat Safe Harbor Agreement~\cite{seal911github,dedaubseal,piperwhitehat}.}

\subsection{Module-scope interventions}
\label{sec:incidents-module}

\paragraph{Delegated Body: Feature-Specific Pauses.}
Aave's role-based control system supports pausing at the pool or reserve level~\cite{AaveACLManager,AaveDPIFreezeProposal}.

\paragraph{Governance Process: Function Toggles.}
MakerDAO's emergency governance proposal during the March 2023 USDC depeg~\cite{MakerDAOEmergency2023} and dYdX's margin adjustments during the YFI incident~\cite{dYdXSushiYFI} exemplify this pattern.

\subsection{Account-scope interventions}
\label{sec:incidents-account}

\paragraph{Signer Set: Key-Based Targeted Restrictions.}
During the November 2025 Balancer exploit, Sonic Labs deployed \texttt{freezeAccount} within two hours~\cite{Sonic2025Freeze}. However, 78.5\% of funds were lost through a \texttt{permit()} bypass. After the March 2022 Ronin Bridge exploit (\$625M), attacker addresses were partially frozen despite 6-day delayed detection~\cite{Ronin2022Exploit}.

\paragraph{Delegated Body: Council-Authorized Remediation.}
In the December 2025 Flow incident, an ``Isolated Recovery'' approach restricted only 1,060 addresses (under 0.01\%)~\cite{FlowPostmortem2026,Flow2025Recovery}. The StakeWise emergency multisig recovered \$20.7M without disrupting stable users~\cite{StakeWise2025Recovery}. Yearn's pxETH burn recovered \$2.4M~\cite{Yearn2025Recovery}.

\paragraph{Governance Process: Stake-Weighted Vote.}
The May 2025 Cetus exploit on Sui produced the most rigorous governance-authorized account intervention: 90.9\% of stake voted ``Yes''~\cite{SuiCetusVote,UnchainedCetus2025}.

\paragraph{The VeChain Classification Dispute.}
VeChain's mechanism is a validator-enforced blocklist authorized by community governance, not a unilateral admin key~\cite{ByBit2025SecurityReport,VeChain2025Refutation}.

\section{Future Directions}
\label{subsec:future}

\begin{itemize}
\item \textbf{Formalizing Heterogeneous Stakeholder Costs.} Different stakeholders bear asymmetric costs from both exploits and interventions.
\item \textbf{Decision Support Tooling.} We have implemented an open-source prototype calculator for real-time mechanism calibration.
\item \textbf{Extended Taxonomy Development.} Pre-execution prevention primitives suggest need for additional timing and enforcement dimensions.
\item \textbf{Cross-Chain Coordination Protocols.} Standardized cross-chain emergency protocols could enable synchronized responses.
\item \textbf{Dynamic Mechanism Selection.} Adaptive approaches based on real-time threat assessment.
\end{itemize}

\fi

\ifdefined\terseversion
\section{Appendix: Governance Coding Notes for the TERSE Revision}
\label{app:governance-coding}

To address the reviewer concern about the Governance success-rate claim, \Cref{tab:governance-coding-terse} lists the Governance-coded cases retained in the 52-case high-fidelity subset after revision, together with the reason each case remains in the Governance bucket. \Cref{tab:governance-reclassified-terse} lists cases that were previously treated too loosely as Governance-adjacent and are now reclassified into their operative authority types.

\begin{table}[h]
\centering
\scriptsize
\begin{tabularx}{\linewidth}{>{\raggedright\arraybackslash}p{2.8cm} >{\raggedright\arraybackslash}p{1.4cm} >{\raggedright\arraybackslash}p{1.2cm} >{\raggedright\arraybackslash}X}
\toprule
\textbf{Case} & \textbf{Scope} & \textbf{Success} & \textbf{Coding rationale} \\
\midrule
Gnosis Chain hard fork & Network & 100.0\% & Retained as Governance because the decisive recovery action was a community-authorized hard fork. Still mixed, because a soft-fork freeze provided earlier containment. \\
Sui/Cetus & Account & 73.0\% & Retained as Governance because the decisive redistribution mandate came from a 90.9\% stake vote. Still hybrid, because validators first froze attacker addresses before the vote. \\
MakerDAO USDC-PSM response & Module & 100.0\% & Retained as Governance because the emergency response was routed through an explicit governance vote rather than a guardian-style admin path. \\
VeChain blocklist & Account & 66.0\% & Retained as Governance because the blocklist authority is described as validator-enforced under community authorization rather than a unilateral admin freeze. \\
Ethereum DAO fork & Network & 100.0\% & Retained as Governance because the decisive intervention was a socially and technically coordinated hard fork authorized at the community level. \\
\bottomrule
\end{tabularx}
\caption{\textbf{Governance-coded cases retained in the TERSE revision.} These five cases form the revised Governance-coded subset used for the paper's constrained success-rate discussion.}
\label{tab:governance-coding-terse}
\end{table}

\begin{table}[h]
\centering
\scriptsize
\begin{tabularx}{\linewidth}{>{\raggedright\arraybackslash}p{3cm} >{\raggedright\arraybackslash}p{2.2cm} >{\raggedright\arraybackslash}X}
\toprule
\textbf{Case} & \textbf{Revised authority} & \textbf{Reason for reclassification} \\
\midrule
Aave v2 guardian pause & Delegated Body & The operative intervention was a Guardian pause, which is better described as bounded delegated authority than as collective governance, even though later governance discussion followed. \\
Alpha Homora V2 / Iron Bank pause & Signer Set & The operative intervention path was core-team-led operational response rather than a collective governance process. \\
\bottomrule
\end{tabularx}
\caption{\textbf{Cases reclassified out of the Governance bucket.} These rows were corrected during the TERSE revision so that the Governance subset better reflects collective authorization rather than guardian/admin intervention.}
\label{tab:governance-reclassified-terse}
\end{table}
\fi

\bibliographystyle{plain}
\bibliography{bib}
\end{document}